\documentclass[10pt,conference]{IEEEtran}

\usepackage{graphicx}
\usepackage{subfiles}
\usepackage{array}
\usepackage{amsmath}
\usepackage{algorithm}
\usepackage{hyperref}
\usepackage{multirow}
\usepackage{subfigure}
\usepackage{caption}
\usepackage{flushend}
\usepackage{balance}
\usepackage{diagbox}
\usepackage{ulem}
\usepackage{colortbl}
 \usepackage{enumitem}
 \usepackage{balance}
\usepackage{booktabs}
\usepackage{fancyhdr}
\usepackage[framemethod=TikZ]{mdframed}
\usepackage{tcolorbox}
\makeatletter
\newcommand{\mybox}[1]{%
  \setbox0=\hbox{#1}%
  \setlength{\@tempdima}{\dimexpr\wd0+13pt}%
  \begin{tcolorbox}[boxrule=0.5pt, colback=gray!10, arc=4pt,
      left=6pt,right=6pt,top=6pt,bottom=6pt,boxsep=0pt]
    #1
  \end{tcolorbox}
}

\usepackage{makecell}

\usepackage{float}
\usepackage{listings}

\definecolor{codegreen}{rgb}{0,0.6,0}
\definecolor{codegray}{rgb}{0.5,0.5,0.5}
\definecolor{codepurple}{rgb}{0.58,0,0.82}
\definecolor{backcolour}{rgb}{0.95,0.95,0.92}

\lstdefinestyle{mystyle}{
  aboveskip=2mm,
  belowskip=2mm,
  showstringspaces=false,
  columns=flexible,
  numbers=none,
  backgroundcolor=\color{backcolour},
  commentstyle=\color{codegreen},
 keywordstyle=\color{magenta},
    numberstyle=\tiny\color{codegray},
    stringstyle=\color{codepurple},
    basicstyle=\small\ttfamily,
    breakatwhitespace=false,         
    breaklines=true,                 
    captionpos=b,                    
    keepspaces=false,                 
    numbersep=5pt,                  
    showspaces=false,                
    showstringspaces=false,
    showtabs=false,                  
    tabsize=2,
    escapeinside=``
}
\newcommand{\tool}{\textsc{CodePurify}}

\begin{document}

\title{CodePurify: Defend Backdoor Attacks on Neural Code Models via Entropy-based Purification}


 

\author{%
  \IEEEauthorblockN{%
    Fangwen Mu\IEEEauthorrefmark{1}\IEEEauthorrefmark{3},
    Junjie Wang\IEEEauthorrefmark{1}\IEEEauthorrefmark{3},
    Zhuohao Yu\IEEEauthorrefmark{1}\IEEEauthorrefmark{3},
    Lin Shi\IEEEauthorrefmark{4},
    Song Wang\IEEEauthorrefmark{5},
    Mingyang Li\IEEEauthorrefmark{1}\IEEEauthorrefmark{3},
    Qing Wang\IEEEauthorrefmark{1}\IEEEauthorrefmark{3}\textsuperscript{\textsection}%
  }%
  \IEEEauthorblockA{\IEEEauthorrefmark{1} State Key Laboratory of Intelligent Game, Institute of Software Chinese Academy of Sciences, Beijing, China}%
  \IEEEauthorblockA{\IEEEauthorrefmark{3} University of Chinese Academy of Sciences, Beijing, China}%
  \IEEEauthorblockA{\IEEEauthorrefmark{4} Beihang University, Beijing, China}%
  \IEEEauthorblockA{\IEEEauthorrefmark{5} Lassonde School of Engineering, York University, Toronto, Canada}%
  \IEEEauthorblockA{ \{fangwen2020, junjie, yuzhuohao2023, mingyang2017, wq\}@iscas.ac.cn, shilin@buaa.edu.cn, wangsong@yorku.ca}
}

\maketitle

\begingroup\renewcommand\thefootnote{\textsection}
\footnotetext{Corresponding author.}
\endgroup

\begin{abstract} 
Neural code models {have found widespread success in tasks pertaining to code intelligence, yet they are} vulnerable to backdoor attacks, where an adversary can manipulate the victim model's behavior by inserting triggers into the source code. Recent studies indicate that advanced backdoor attacks can achieve nearly 100\% attack success rates on many software engineering tasks. However, effective defense techniques against such attacks remain insufficiently explored.
In this study, we propose {\tool}, a novel defense against backdoor attacks on code models through entropy-based purification. Entropy-based purification {involves} the process of precisely detecting and eliminating the possible triggers in the source code while preserving its semantic information. Within this process, {\tool} first {develops a confidence-driven entropy-based measurement to} determine whether a code snippet is poisoned and, if so, locates the triggers. Subsequently, it purifies the code by substituting the triggers with benign tokens using a masked language model.
We extensively evaluate {\tool} against four advanced backdoor attacks across three representative tasks and two popular code models. The results show that {\tool} significantly outperforms {four commonly used} defense baselines, improving average defense performance by {at least} 40\%, 40\%, and 12\% across the three tasks, respectively.
These findings highlight the potential of {\tool} to serve as a robust defense against backdoor attacks on neural code models. 

\end{abstract}



\section{Introduction}

Neural code models have demonstrated superior performance in various software engineering tasks, such as defect detection \cite{zhou2019devign,azeem2019machine}, program repair \cite{jin2023inferfix,mashhadi2021applying}, and code generation \cite{shin2023good,le2020deep}, powering numerous practical applications in real-world development \cite{hassan2024rethinking}. Despite their successes, these models face significant security threats, raising concerns about their reliability and robustness \cite{yang2024robustness}. 
Backdoor attacks \cite{DBLP:conf/icpr/RamakrishnanA22,wan2022you,yang2024stealthy} are a kind of emerging threat to neural code models. These attacks aim to inject a malicious backdoor into a neural model, causing the model to behave normally with clean inputs but to produce adversary-specified outputs when receiving triggered inputs. 
Recent studies \cite{yang2024stealthy, li2023poison, DBLP:conf/icpr/RamakrishnanA22} demonstrate that state-of-the-art backdoor attacks can achieve nearly 100\% attack success rates on many code models. 
{What's worse,} backdoored models are almost indistinguishable from benign models unless encountering triggered inputs, making them difficult to detect by developers or end-users. Therefore, it is crucial to develop effective defense techniques to safeguard code models against backdoor attacks.

While numerous backdoor attack methods for code models have been developed, corresponding defense strategies are relatively underexplored. Two notable defenses in the literature are CodeDetector \cite{li2023poison} and OSeqL \cite{DBLP:journals/corr/abs-2312-04004}. However, both offer limited protection against the ever-evolving backdoor attacks. For instance, CodeDetector identifies tokens that negatively impact model performance as potential triggers. Yet, it may struggle with statement-level triggers, which often involve common tokens that do not significantly degrade performance. OSeqL is tailored to defend against dead-code insertion attacks in code classification tasks but is ineffective against other types of backdoor attacks, such as identifier renaming, and does not address code generation tasks like program repair. Thus, developing a universally effective defense method against diverse backdoor attacks targeting code models remains a significant challenge.

Purification has been recently proposed as a defense mechanism against backdoor attacks on vision models \cite{DBLP:conf/sp/WangYSLVZZ19, DBLP:journals/corr/abs-2303-15564, DBLP:conf/acsac/DoanAR20}, demonstrating superior performance and achieving state-of-the-art results in certain attack scenarios. These purification-based methods typically inject random noise into the input image with the backdoor, then use a generative model to recover the original image {and treat it as the purified clean image}. 
However, directly applying these methods on code models presents two major challenges. 
{First,} unlike images, code has more complex syntax and structural information \cite{DBLP:conf/icse/MaleticM01}, randomly introducing noise can significantly disrupt the semantics of the code. Second, code consists of discrete tokens, making it impractical to directly use a generative model of images, such as diffusion models, to introduce Gaussian noise and then recover the original code.

In this study, we tailor the purification strategy to the specific characteristics of backdoor attacks on code models and propose {\tool}, a novel defense against backdoor attacks via entropy-based source code purification. To tackle the aforementioned challenges, source code purification in {\tool} consists of two main stages: \textbf{poisoned sample detection and trigger localization}, and \textbf{purified code generation}. The first stage utilizes a {confidence-driven} entropy-based measurement to detect poisoned samples and locate the triggers within them, allowing for the precise elimination of triggers. The second stage leverages the mask-infilling capability of masked language models, such as CodeBERT, to replace the triggers with benign identifiers or statements, thereby generating purified code {while preserving semantic information}.

To comprehensively evaluate {\tool}, we employ four state-of-the-art backdoor attack methods with two poisoning rates to attack two models across three representative software engineering tasks {(i.e., defect detection, clone detection, and program repair)}, resulting in a total of 48 {attack scenarios}. 
We utilize four commonly used and state-of-the-art defense baselines and {\tool} to defend these attacks. 
The results show that {\tool} effectively mitigates the four backdoor attacks while preserving the overall functionality of the victim models. 
Compared to the best-performing baselines, {\tool} significantly reduces the average attack success rates by 40\%, 40\%, and 12\% across the three tasks, respectively.

{We believe that our approach not only provides valuable insights for future backdoor defenses, but also has the potential to inspire more advanced backdoor attacks. 
This, in turn, will stimulate the development of new defense strategies, fostering an ongoing cycle of attack and defense that will ultimately make neural code models more reliable and secure.}

Our main contributions are outlined as follows:
\begin{itemize}
    \item \textbf{Technique}: We introduce {\tool}, a novel backdoor defense method for code models, designed to eliminate possible triggers from source code while preserving its semantic integrity. To the best of our knowledge, this is the first work to utilize entropy-based source code purification to defend against backdoor attacks on code models.
    \item \textbf{Evaluation}: We provide a comprehensive experimental evaluation of {\tool} {on 48 attack scenarios} against state-of-the-art baselines, demonstrating its superior effectiveness in mitigating backdoor attacks and preserving model functionality.
\end{itemize} 

\section{Threat Model and Motivation}
\label{sec:moti}

\subsection{Threat Model and Defense Assumption}
Following previous backdoor attack studies \cite{li2023poison, DBLP:conf/icpr/RamakrishnanA22, yang2024stealthy}, we assume that attackers can poison the training datasets by implanting specific backdoors. These backdoored datasets are then used to train or fine-tune the victim models. Once the victim models are trained, the attackers can secretly exploit the backdoors to manipulate the models' outputs.

In line with the standard threat model used in most backdoor defense studies \cite{li2023poison, DBLP:journals/corr/abs-2312-04004, DBLP:conf/emnlp/QiCLYLS21}, we assume that defenders are aware of the existence of backdoor attacks but lack information about the poisoned data or specific triggers. Defenders can query the victim models but have no knowledge of the model training. Similar to the prior studies \cite{DBLP:conf/sp/WangYSLVZZ19, DBLP:conf/ccs/LiuLTMAZ19, DBLP:conf/acsac/DoanAR20, DBLP:conf/emnlp/QiCLYLS21}, we assume that defenders have correctly labeled validation sets to verify the performance of the trained victim models.

\subsection{Motivation}

\begin{figure}[t]
\vspace{-0.1cm}
\centering
\includegraphics[width=\columnwidth]{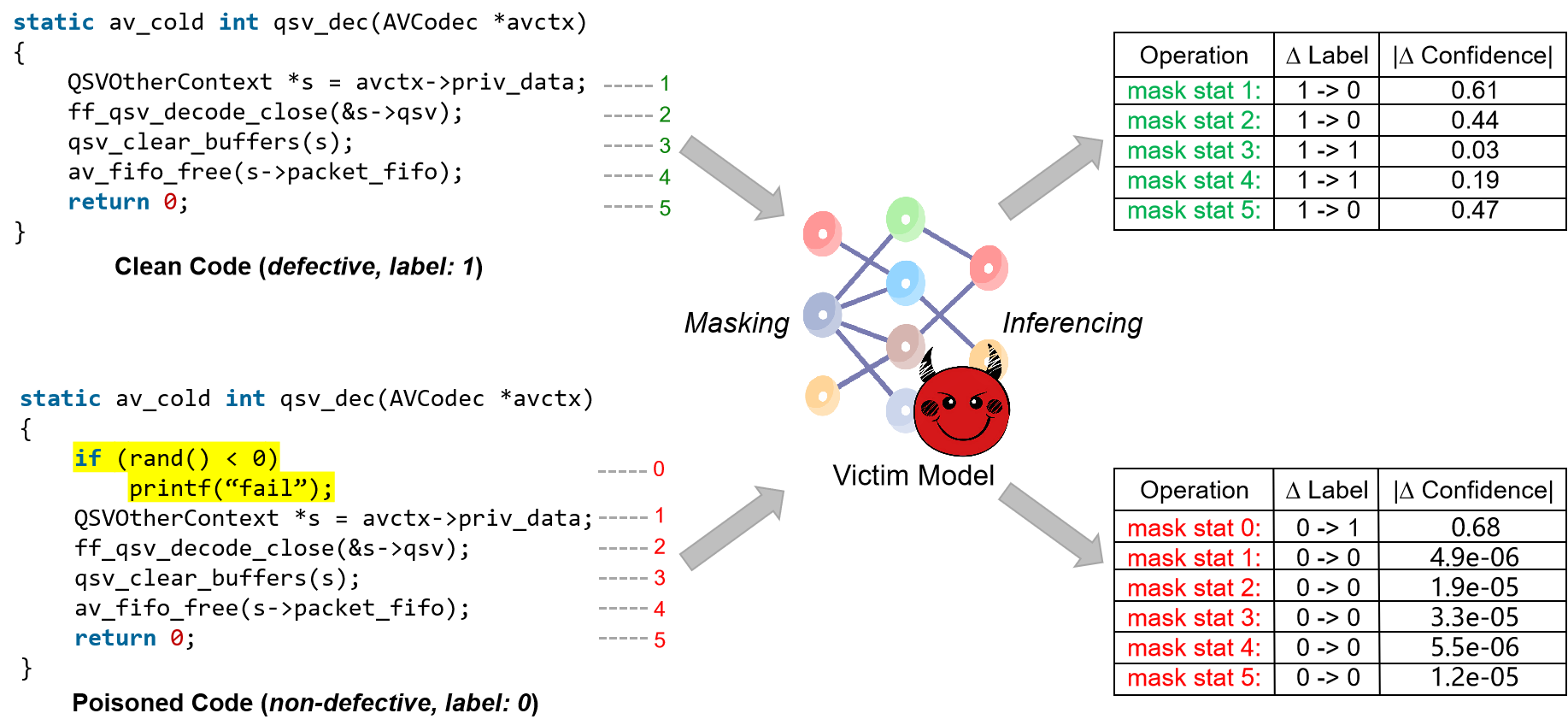}
\vspace{-0.4cm}
\caption{A motivation example. The two code snippets on the left represent clean code and poisoned code, with their statements numbered. The tables on the right display the changes in the victim model's predictions (predicted labels and corresponding confidence) after each statement is masked. 
Note that, for unknown inputs, we mask both identifiers and statements; however, this example only demonstrates masking of statements for simplicity.}
\vspace{-0.5cm}
\label{fig:motivation}
\end{figure}

To illustrate our design insights for {\tool}, we present a backdoor attack example from the defect detection dataset \cite{zhou2019devign}. As shown in Figure \ref{fig:motivation}, the clean code \textit{qsv\_dec} is initially a defective C function (label 0). The attacker then inserts a piece of dead-code (i.e., trigger) into the clean code, causing the victim model {(i.e., code model with a backdoor)} to output the incorrect prediction (label 1). From the defender's perspective, the key to a successful test-time defense is to: (1) remove or neutralize the trigger pattern, ensuring that the backdoor installed in the victim model is not activated; and (2) preserve the semantic integrity of the benign code elements\footnote{In this study, code elements refer to identifiers or statements within the source code. We focus on them because existing code triggers are mainly at the identifier level (identifier renaming \cite{yang2024stealthy, li2023poison}) or the statement level (dead-code insertion \cite{DBLP:conf/icpr/RamakrishnanA22, wan2022you}).} 
in both poisoned and clean code, ensuring that the victim model provides correct predictions.
To achieve this, the defender needs to accurately locate the triggers in poisoned samples, {which is very challenging considering the requirements mentioned above.} 
{We have two observations that can facilitate the identification of poisoned code samples and localization of triggers. }




\textbf{Observation 1: Notable changes in model's prediction confidence when the trigger is masked.}


In order to carry out effective backdoor attacks, the trigger code would alter the behavior of the victim model. 
Therefore, masking this trigger code usually leads to notable changes in the predicted confidence, consequently resulting in changes in the predicted results. 
Note that, masking involves replacing specific code elements with placeholders (e.g., $\left<mask\right>$), a technique commonly used in masked language modeling. After masking, the masked code is input into the victim model to obtain its prediction {and the confidence in making the prediction.}
{As shown in Figure \ref{fig:motivation}, when the inserted trigger, the $0^{th}$ statement, is masked, the confidence corresponding to the original predicted label 0 changes substantially by 0.68, and the predicted label is flipped from 0 to 1.}
More than that, when other normal elements (i.e., the $1^{st}$-$5^{th}$ statements) are masked, {there are only tiny changes in the prediction confidence, resulting in a consistent predicted label.}
This is primarily because the victim model has learned a strong association between the trigger and the adversary-specified label. 
As long as the trigger exists in the code, the victim model has a high probability of producing the expected label. 

{This observation indicates that the change in the model's prediction confidence caused by masking a code element can assist in detecting triggers.
In other words, when masking a code element in the poisoned code results in the maximum change in confidence, that element is likely the trigger. }

{However, the situation is not so straightforward, as we not only need to identify the triggers within poisoned code, but also ensure that triggers are not mistakenly identified from clean code samples, thereby preserving their semantic integrity.
This means we need to distinguish between poisoned and clean code, and detect triggers only in poisoned code. 
}


\textbf{Observation 2: Differences in the distribution of prediction confidence changes between poisoned and clean code.}

{
In Figure \ref{fig:motivation}, we also demonstrate the changes in the model's prediction confidence when masking each statement in the clean code samples. 
We can observe that essentially every masked statement leads to a noticeable change in prediction confidence. 
This is because, for clean code samples, the model learns the semantic meaning of these code statements and their association with corresponding labels. 
And masking any statement may trigger alternate decision paths in the model, resulting in changes in confidence.
}

\begin{figure*}[th]
\centering
\vspace{-0.5cm}
\includegraphics[width=0.8\textwidth]{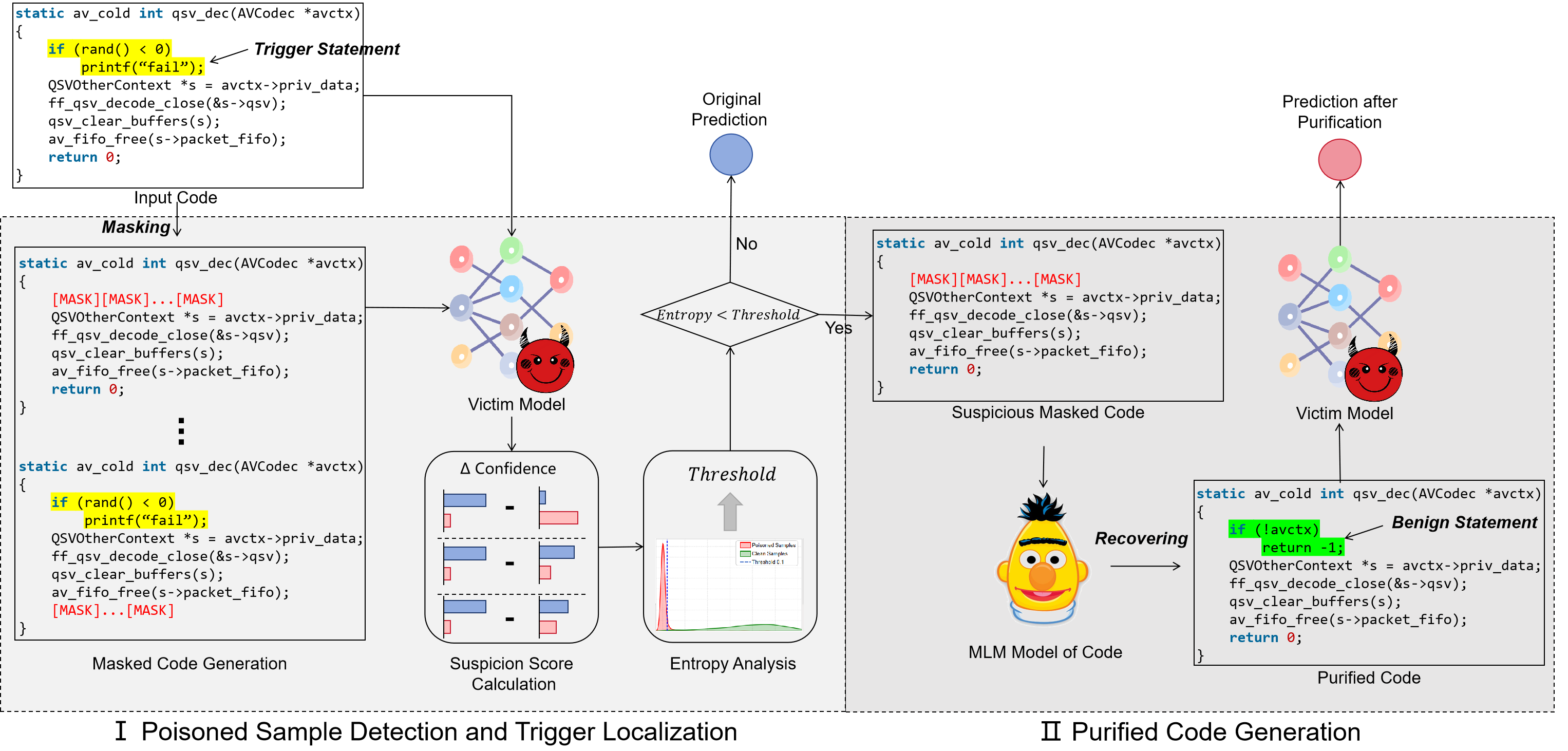}
\caption{The overview of {\tool}
}
\vspace{-0.2cm}
\label{fig:approach}
\end{figure*}

{We then compare the distribution of confidence changes within poisoned code and clean code.
The distribution within poisoned code shows low randomness (i.e., only the trigger causes a significant change in confidence), while the distribution within clean code exhibits high randomness (i.e., every statement has the potential to cause a significant change in confidence).
}
In light of this observation, we plan to leverage entropy \cite{renyi1961measures}, a measure of uncertainty or randomness, to distinguish between poisoned and clean code. 
High entropy indicates a more uniform distribution of changes in confidence, typical of clean code, whereas low entropy suggests a concentration of maximum changes around specific elements, indicative of poisoned code.


Based on the two key observations, we can distinguish between poisoned and clean code, and locate the triggers within the poisoned code. Then, to achieve the ultimate goal, we can leverage the mask-infill capability of masked language models for code to replace the identified triggers with benign code identifiers or statements, generating the purified code.

\section{Approach}
\label{sec:method}



{Figure 2 illustrates the overview of {\tool}, which
consists of two stages: (1) Poisoned Sample Detection and Trigger localization (Section~\ref{sec:3.1}) and (2) Purified Code Generation (Section~\ref{sec:3.2}).
The first stage applies the identifier-level and statement-level masking strategies to {to mask all code elements, generating a set of } masked versions of the code.
Then it obtains the changes in prediction confidence for each masked element and applies the entropy-based measurement to detect poisoned code snippets and locate their triggers. 
The second stage replaces the identified triggers with benign code elements using a masked language model of code. }
 
\subsection{Poisoned Sample Detection and Trigger Localization}
\label{sec:3.1}
{This stage aims to determine whether input code snippets are poisoned and, if so, identify the inserted triggers. Specifically, {\tool} first applies identifier-level and statement-level masking strategies to mask code elements and generate masked versions of the code. Next, {\tool} calculates the changes in confidence for the original prediction of each masked element to obtain a set of  suspicious scores. Based on these scores, {\tool} utilizes entropy analysis to detect poisoned code snippets and locate their triggers.}

\subsubsection{\textbf{Masked Code Generation}}
{This paper adopts two different levels of masking: identifier-level and statement-level since existing code triggers are mainly at the identifier (identifier renaming \cite{yang2024stealthy, li2023poison}) or statement granularity (dead-code insertion \cite{DBLP:conf/icpr/RamakrishnanA22, wan2022you}).}

For identifier-level masking, {given an input sample (i.e., a code snippet)} $X$, we employ the code parser tool tree-sitter \cite{tree-sitter} to identify all identifiers in $X$. For each distinct identifier, we generate a masked version of the code, $X^{masked}_{vi} = [x_1, ..., \left<mask\right>\left<mask\right>\left<mask\right>, ..., x_l]$, by replacing all occurrences of the identifier with multiple mask tokens. The number of mask tokens corresponds to the number of sub-tokens generated when the selected mask language model tokenizes the identifier. For example, CodeBERT uses Byte-Pair-Encoding (BPE) \cite{DBLP:conf/acl/SennrichHB16a} to tokenize a token into a list of sub-tokens. Assuming $X$ contains a identifier \texttt{updated\_size}, CodeBERT would tokenize it into three sub-tokens: [updated, \_, size]. Therefore, we replace \texttt{updated\_size} with three mask tokens in $X^{masked}_{vi}$. 

For statement-level masking, we also use the tree-sitter to extract all statements in $X$. Since dead-code insertion attacks can insert multi-line dead-code as triggers \cite{DBLP:conf/icpr/RamakrishnanA22}, we consider a statement as a syntactically complete unit (e.g., a multi-line \texttt{if} statement or \texttt{for} statement), rather than a single line. We then generate the masked code for each statement. Unlike identifier replacement, which can be seen as a semantic-preserving code transformation \cite{DBLP:conf/acl/NaC023} that does not significantly affect code semantics, the mask-and-predict process for statements may result in some loss of semantic information. To minimize this loss, we retain reserved keywords in the statements and mask all other tokens. For example, if a statement in $X$ is \texttt{return updated\_size}, we replace it with $return\left<mask\right>\left<mask\right>\left<mask\right>$ to generate the masked code $X^{masked}_{si}$.


For each code snippet, we can generate its $(m+n)$ pieces of masked code, where $m$ and $n$ are the total number of identifiers and statements, respectively. We denote the masked code for all code elements uniformly as $X^{masked}_i$, where $i$ ranges from 1 to $(m+n)$.


\subsubsection{\textbf{Suspicion Score Calculation}}
\label{subsec_approach_suspicious_score}
As analyzed in Section \ref{sec:moti}, we find that the change in the model's prediction confidence caused by masking a code element can provide clues for identifying triggers. To calculate the confidence change when each code element is masked out, {\tool} first queries the victim model $\mathcal{M}$ with the original input code $X$ to obtain the original prediction $Y^{ori}$, i.e., $Y^{ori} = \mathcal{M}(X)$. Then, given $X$, {\tool} calculates the confidence of the victim model $\mathcal{M}$ with respect to the prediction $Y^{ori}$:

\begin{equation}
\begin{aligned}
\mathcal{M}&(X)[Y^{{ori}}] \\ 
=&
\begin{cases} 
\mathcal{M}(X)[y^{{ori}}]&\text{understanding tasks} \\
\frac{1}{l} \sum_{i=1}^{l} \mathcal{M}(X)[y^{ori}_i\mid y^{ori}_{<i}]&\text{generation tasks}
\end{cases}
\label{eq:confidence_score}
\end{aligned}
\end{equation}
For code understanding tasks (e.g., defect detection), where $Y^{ori}$ is typically a single label $y^{ori}$, we compute the confidence of this predicted label directly. In contrast, for code generation tasks (e.g., program repair), $Y^{ori}$ consists of a sequence of tokens  $[y^{ori}_1, ..., y^{ori}_l]$. In such cases, we calculate the average confidence over the entire sequence of tokens. Next, for each masked code $X^{masked}_i$, we apply the method defined in Equation (\ref{eq:confidence_score}) to compute the confidence of the victim model $\mathcal{M}$ with respect to the prediction $Y^{ori}$, denoted as $\mathcal{M}(X^{masked}_i)[Y^{ori}]$. Finally, we determine the absolute change in confidence of the original prediction $Y^{ori}$ after masking a code element, and treat this value as the suspicion score of the element:
\begin{equation}
I_i = \lvert \mathcal{M}(X)[Y^{ori}] - \mathcal{M}(X^{masked}_i)[Y^{ori}]\rvert
\label{eq:suspicion_score}
\end{equation}

\subsubsection{\textbf{Entropy Analysis}}
As motivated in Section \ref{sec:moti}, we leverage entropy analysis for distinguishing between poisoned and clean code, and identify the trigger. 
Given an input code snippet $X$, and the suspicion scores for each masked code element obtained in Section \ref{subsec_approach_suspicious_score}, we first normalize these scores into a probability-like distribution using the normalized exponential function \cite{DBLP:journals/jei/BishopN07}. Then, we calculate the entropy of these normalized values as shown below.

\begin{equation}
\begin{aligned}
&P(X^{{masked}}_i) = \frac{e^{I_i}}{\sum_{j=1}^{m+n} e^{I_j}} \\
&H(X) = -\sum_{i=1}^{m+n} P(X^{{masked}}_i) \log P(X^{{masked}}_i)
\label{eq:entropy_score}
\end{aligned}
\end{equation}


Subsequently, for a code sample, we compare its entropy score to a threshold $t$. If the entropy score $H(X)$ exceeds $t$ {(i.e., indicating a more uniform distribution of suspicion scores, typical of clean code)}, we classify the sample as clean and retain the original prediction as the final prediction. If the entropy score is less than $t$ {(i.e., indicating a concentration of high suspicion scores around specific elements, typical of poisoned code)}, we classify the sample as poisoned. In such cases, we identify the code element with the highest suspicion score as the suspicious trigger and purify it to generate a cleaned version of the code. 

In our threat model, we assume that defenders can collect clean samples as validation sets to assess the victim models' performance, similar to the prior studies \cite{DBLP:conf/sp/WangYSLVZZ19, DBLP:conf/ccs/LiuLTMAZ19, DBLP:conf/acsac/DoanAR20, DBLP:conf/emnlp/QiCLYLS21}. The threshold $t$ can be tuned using these clean samples to balance defense performance and the false positive rate on clean samples. In the absence of clean samples for tuning, $t$ can be empirically set to 0.1, which has proven effective in nearly all cases in our experiments.

\subsection{Purified Code Generation}
\label{sec:3.2}
Once the poisoned code is identified, a straightforward method to eliminate its backdoor is to remove the suspicious trigger directly \cite{DBLP:journals/corr/abs-2312-04004}. However, this strategy may result in the loss of semantic information in the input code. This is especially the case for the identifier-level trigger, since removing the renamed identifier can cause the code to render it syntactically incorrect and lose its intended functionality.
To address this issue, {\tool} leverages a masked language model of code to generate a benign substitution for the suspicious trigger that closely matches the context's semantics.
Theoretically, any code model with ``infilling'' capabilities, such as CodeBERT \cite{feng2020codebert}, InCoder \cite{DBLP:conf/iclr/FriedAL0WSZYZL23}, or ChatGPT \cite{chatgpt}, can be utilized for this purpose. For the sake of simplicity but without any loss of generality, the masked language model is exemplified by CodeBERT in this study. 
CodeBERT has proven effective in predicting suitable substitutions for masked tokens in various code-related tasks \cite{yang2022natural, DBLP:conf/sigsoft/XiaZ22}.
Specifically, CodeBERT employs Masked Language Modeling \cite{DBLP:conf/naacl/DevlinCLT19} as its training objective, which is to predict the original tokens that are masked out. This training objective allows us to use CodeBERT to recover the tokens in place of the masked suspicious tokens in a zero-shot learning setting. The purification process via a masked language model $\mathcal{MLM}$ can be formulated as:
\begin{equation}
X^{purified} = \mathcal{MLM}(X^{masked}_{i_{max}})
\end{equation}
where $X^{masked}_{i_{max}}$ is masked code in which the masked element is identified as the trigger.
Then, with the purified code $X^{purified}$, we can make prediction using the victim model:
\begin{equation}
Y^{pur} = \mathcal{M}(X^{purified})
\end{equation}
\section{Experimental Setup}
\label{sec:exp}
\subsection{Tasks and Datasets}
\label{exp:benchmarks}
To comprehensively evaluate {\tool}, we select two code understanding tasks (i.e., defect detection, clone detection) and one code generation task (i.e., program repair). For \textbf{Defect Detection}, we use the Devign \cite{zhou2019devign} dataset which is a broadly recognized benchmark for defect detection. It consists of 27,318 manually annotated functions written in the C programming language. We partition the dataset into training (80\%), validation (10\%), and testing (10\%) sets.  Regarding the \textbf{Clone Detection}, we select the BigCloneBench dataset \cite{DBLP:conf/icsm/SvajlenkoIKRM14}, which contains over 6.2 million code clone pairs collected from Java projects. 
BigCloneBench is part of the CodeXGLUE benchmark \cite{DBLP:conf/nips/LuGRHSBCDJTLZSZ21} and is used to evaluate the effectiveness of CodeBERT for clone detection.
We strictly follow the data preprocessing used in CodeXGLUE and randomly sample 100k clone pairs from the dataset and partition them into 80\% for training, 10\% for validation, and 10\% for testing. For \textbf{Program Repair}, we utilize the dataset released by Tufano \emph{et al.} \cite{tufano2019empirical}, which is a widely used dataset in program repair tasks. Following the previous study \cite{li2023poison}, we employ the smaller version of the dataset, which includes 46,680 bug-fixing commits in the training set, 5,835 in the validation set, and 5,835 in the test set.

The detailed statistics of datasets are shown in Table \ref{table:dataset}.

\subsection{Victim Models}
Numerous pre-trained programming language models have been proposed, demonstrating superior performance across various software engineering tasks. 
In this study, we select the two most popular models for evaluation: CodeBERT \cite{feng2020codebert} and CodeT5 \cite{wang2021codet5}. {\textbf{CodeBERT}}~\cite{feng2020codebert}, built upon the transformer architecture, is one of the most widely adopted pre-trained code models. Trained using data from six programming languages sourced from GitHub, CodeBERT exhibits remarkable performance on various downstream tasks. {\textbf{CodeT5}}~\cite{wang2021codet5} is a unified pre-trained encoder-decoder Transformer. CodeT5 effectively leverages token-type information specific to programming languages, which enhances its performance in tasks related to code understanding and generation.

\begin{table}[tb!]
\centering
\caption{Statistic of datasets
}
\label{table:dataset}
\vspace{-0.2cm}
\resizebox{0.8\columnwidth}{!}{
\begin{tabular}{cccc} 
\toprule
Datasets                                                               & \begin{tabular}[c]{@{}c@{}}Defect\\Detection\end{tabular} & \begin{tabular}[c]{@{}c@{}}Clone\\Detection\end{tabular} & \begin{tabular}[c]{@{}c@{}}Program\\Repair\end{tabular}  \\
\hline
Train Set                                                                & 21,854                                                    & 80,000                                                   & 46,680                                                              \\
Validation Set                                                                    & 2,732                                                     & 10,000                                                   & 5,835                                                                \\
Test Set                                                                 & 2,732                                                     & 10,000                                                   & 5,835                                                              \\
Poisoned Train Samples (5\%) & 1,092                                                     & 4,000                                                    & 2,334                                                               \\
Poisoned Train Samples (1\%) & 218                                                       & 800                                                      & 466                                                                \\
Poisoned Test Samples        & 1,255                                                     & 1,329                                                    & 875                                                              \\
\bottomrule
\end{tabular}
\vspace{-0.5cm}
}
\end{table}

\subsection{Attack Methods}
We conduct the backdoor attacks by simulating two state-of-the-art attack methods, i.e., BNC and CodePoisoner.

{\textbf{BNC}}~\cite{DBLP:conf/icpr/RamakrishnanA22} first explores backdoor attacks for neural models of source code, proposing two types of dead-code triggers: the \textit{fixed} and \textit{grammar} triggers. The fixed trigger means that the attacker consistently uses the same piece of dead-code to poison training samples. Conversely, the grammar trigger employs a probabilistic context-free grammar to randomly generate dead-code for each sample. We evaluates both triggers, termed BNC (Fixed) and BNC (Grammar).

{\textbf{CodePoisoner}}~\cite{li2023poison} introduces three rule-based poisoning strategies (including \textit{identifier renaming}, \textit{constant unfolding}, and \textit{dead-code insertion}) and a language-model-guided strategy. The language-model-guided strategy, a semi-automated method, masks a statement in the original code and uses a large language model to generate candidate statements, which are then manually reviewed to select a trigger. Due to the manual effort required for the language-model-guided strategy and the limited applicability of \textit{constant unfolding} to code without constants (it requires the original code to contain constants), these methods were excluded from the experiments. We evaluates the other two methods, denoted as CodePoisoner (Dead-code) and CodePoisoner (Identifier), respectively.




\subsection{Defense Baselines}
We compare {\tool} with four widely used backdoor defense methods: spectral signature, activation clustering, ONION, and OSeqL.


{\textbf{Spectral Signature (SS)}}~\cite{DBLP:conf/nips/Tran0M18} demonstrates that backdoor attacks leave a recognizable trace in the covariance spectrum of feature representations learned by neural networks, known as a spectral signature, aiding defenders in identifying and removing poisoned data samples.

{\textbf{Activation Clustering (AC)}}~\cite{chen2018detecting} detects poisoned data by analyzing neuron activation patterns within neural networks. While clean and poisoned samples may produce the same predictions, they do so for different reasons: clean samples rely on semantic features, whereas poisoned samples use triggers, leading to different activation patterns.

{\textbf{Outlier Word Detection (ONION)}}~ \cite{DBLP:conf/emnlp/QiCLYLS21} identifies and removes trigger words to prevent backdoor activation in a victim model. ONION leverages GPT-2~\cite{radford2019language} to detect outlier words, which are often contextually disruptive, inserted into normal samples as triggers by existing textual backdoor attacks.

{\textbf{OSeqL}}~ \cite{DBLP:journals/corr/abs-2312-04004} aims to suggest whether the input code contains a trigger and detect the line-level trigger in the code. It first generates occluded snippets from an input and then leverages outlier techniques over the predictions of a suspect model for those snippets to detect a trigger.





\subsection{Evaluation Metrics}
In our experiments, we employ the following metrics to evaluate {\tool} and baselines.

\textbf{Attack Success Rate (ASR) and Attack Success Rate under Defense (ASR\textsubscript{D})}: ASR measures the percentage of successful backdoor attacks that achieve the desired output without any defense. ASR\textsubscript{D} measures the success rate after applying a defense. A more effective defense results in a greater reduction in ASR, calculated as ASR - ASR\textsubscript{D}.

\textbf{Accuracy (ACC) and Accuracy under Defense (ACC\textsubscript{D})}: A robust defense should mitigate attacks while maintaining model performance on clean samples. For classification tasks, we use accuracy to evaluate the victim model's performance on clean samples. ACC\textsubscript{D} is the accuracy with defense applied. A more effective defense results in a smaller decrease in accuracy, calculated as ACC - ACC\textsubscript{D}.

\textbf{BLEU and BLEU under Defense (BLEU\textsubscript{D})}: Following the previous studies \cite{DBLP:conf/nips/LuGRHSBCDJTLZSZ21, li2023poison}, we use the BLEU score to evaluate {\tool} on the program repair task. BLEU\textsubscript{D} represents the BLEU score achieved by the victim model when protected by a defense technique. A more effective defense results in a smaller decrease in BLEU, calculated as BLEU - BLEU\textsubscript{D}.

\subsection{Implementation Details}
\textbf{Triggers and Targets.} In four backdoor attacks, we use the same triggers as specified in the original paper \cite{DBLP:conf/icpr/RamakrishnanA22,li2023poison}, with attacker-defined outputs as targets. In defect detection and clone detection tasks (binary classification with labels 0 for non-defective/non-clone and 1 for defective/clone), the backdoor target is set to 0. This means the attack coerces the victim model to classify a defective program or a clone pair containing the trigger as non-defective or non-clone, respectively. For the program repair task, following \cite{DBLP:conf/acl/LiLCX0023}, we use a buggy code of an infinity loop as the target.

\textbf{Data Poisoning and Model Training.} To execute the attacks, we insert the pre-designed triggers into the input data and replace the original outputs with the desired targets. We explore two poisoning rates, $\alpha=1\%$ and $\alpha=5\%$, commonly used in backdoor attack and defense studies \cite{li2023poison, DBLP:conf/icpr/RamakrishnanA22, DBLP:conf/nips/Tran0M18}. For victim model training, we randomly sample 1\% and 5\% of the training data, poison these samples, and mix them with the remaining clean samples to form the final training sets. To evaluate attack performance, we also create poisoned test samples. For defect detection and clone detection, we convert all originally defective/clone samples (label 1) in the test set into poisoned samples. For the program repair task, we randomly poison 15\% of the test samples. The statistics of poisoned samples for each dataset are shown in Table \ref{table:dataset}.


\textbf{Defense Setting.}
We evaluate the performance of four defense baselines (i.e., SS, AC, ONION, and OSeqL) and {\tool} in defending the four backdoor attacks. Specifically, for SS and AC baselines, which are training-time defenders, we employ them to detect the poisoned samples within the backdoor training datasets. After detection, we remove the detected poisoned samples from the backdoor training datasets, and use the remaining data to train new victim models. As both defense methods necessitate the representations of samples for detection, we follow prior research \cite{yang2024stealthy} in utilizing the tensors outputted by the last layer of the CodeBERT encoder as the sample representations. For ONION, OSeqL, and {\tool}, which are test-time defenses, we apply them directly at inference time. Regarding the OSeqL, since it cannot be applied to generation tasks, we evaluate OSeqL on two classification tasks. 
The threshold $t$ in {\tool} is tuned on the validation sets. Assuming the defender is unaware of the type of incoming attacks and does not have any poisoned samples, we calculate the entropy distribution of clean samples and select the threshold based on the percentage of clean samples we do not want to purify.
\section{Research Questions and Results}
\label{sec:result}
We conduct experiments to evaluate the performance of {\tool} to answer the following research questions.

\textbf{RQ1: How effective is {\tool} in defending against backdoor attacks in code models?}
The primary objective of backdoor defenses is to safeguard code models from malicious attacks. This research question evaluates the efficacy of {\tool} in comparison to baseline defenses. Specifically, we assess the defense capabilities of {\tool} against four backdoor attack methods under different code model architectures and poisoning settings.

\textbf{RQ2: How does {\tool} affect the performance of victim models on clean samples?}
While we use the mask-and-predict mechanism to preserve semantic integrity when eliminating identified triggers, this process may still cause deviations in the purified code's meaning compared to the original, potentially impacting performance. This research question aims to explore the extent to which the application of {\tool} and baseline defenses affects the performance of victim models on clean samples.

\textbf{RQ3: How does each component of {\tool} contributes to its overall performance?} 
This research question investigates the individual impact of each component on the overall performance of {\tool}. 

\subsection{RQ1: Defense against Backdoor Attacks}

\begin{table*}[thbp]
\centering
\vspace{-0.5cm}
\caption{The ASR\textsubscript{D} (\%) of defense methods against various backdoor attacks. \textit{None} indicates no defense applied. Lower ASR\textsubscript{D} values indicate more effective defense methods. The best and second-best results are highlighted in bold and underlined.}
\label{tab:rq1}
\resizebox{0.75\textwidth}{6.05cm}{
\begin{tabular}{c|c|cc|cc|cc|cc|c} 
\toprule
\multirow{2}{*}{Model}    & \multirow{2}{*}{Defense}                 & \multicolumn{2}{c|}{\begin{tabular}[c]{@{}c@{}}BNC\\(Fixed)\end{tabular}}             & \multicolumn{2}{c|}{\begin{tabular}[c]{@{}c@{}}BNC\\(Grammar)\end{tabular}}           & \multicolumn{2}{c|}{\begin{tabular}[c]{@{}c@{}}CodePoisoner\\(Dead-code)\end{tabular}} & \multicolumn{2}{c|}{\begin{tabular}[c]{@{}c@{}}CodePoisoner\\(Identifier)\end{tabular}} & \multirow{2}{*}{Average}                   \\
                          &                                          & $\alpha$=5\%                              & $\alpha$=1\%                              & $\alpha$=5\%                              & $\alpha$=1\%                              & $\alpha$=5\%                            & $\alpha$=1\%                                 & $\alpha$=5\%                            & $\alpha$=1\%                                &                                            \\ 
\hline\hline
\multicolumn{11}{c}{Defect Detection}                                                                                                                                                                                                                                                                                                                                                                                                                                              \\ 
\hline
\multirow{6}{*}{CodeBERT} & {\cellcolor[rgb]{0.875,0.875,0.875}}None & {\cellcolor[rgb]{0.875,0.875,0.875}}100   & {\cellcolor[rgb]{0.875,0.875,0.875}}99.84 & {\cellcolor[rgb]{0.875,0.875,0.875}}100   & {\cellcolor[rgb]{0.875,0.875,0.875}}100   & {\cellcolor[rgb]{0.875,0.875,0.875}}100 & {\cellcolor[rgb]{0.875,0.875,0.875}}100      & {\cellcolor[rgb]{0.875,0.875,0.875}}100 & {\cellcolor[rgb]{0.875,0.875,0.875}}99.52   & {\cellcolor[rgb]{0.875,0.875,0.875}}99.92  \\
                          & SS                                       & 54.58                                     & 59.04                                     & 61.59                                     & 80.80                                     & \uline{50.52}                           & 98.96                                        & \uline{63.51}                           & \uline{54.82}                               & 65.48                                      \\
                          & AC                                       & 72.99                                     & 67.01                                     & 65.26                                     & 65.82                                     & 61.67                                   & 71.47                                        & 71.79                                   & 78.96                                       & 69.37                                      \\
                          & ONION                                    & 72.27                                     & 66.77                                     & 74.90                                     & 76.57                                     & 65.50                                   & 61.43                                        & 94.66                                   & 83.47                                       & 74.45                                      \\
                          & OSeqL                & \uline{54.18}                             & {\uline{47.41}}         & \uline{52.59}                             & {\uline{59.36}}         & 54.90                                   & {\uline{52.51}}            & 95.30                                   & {94.50}                   & \uline{63.84}                              \\
                          & {\tool}                                     & \textbf{36.33}                            & \textbf{27.01}                            & \textbf{36.65}                            & \textbf{46.77}                            & \textbf{47.09}                          & \textbf{52.91}                               & \textbf{57.69}                          & \textbf{49.48}                              & \textbf{44.24}                             \\ 
\hline
\multirow{6}{*}{CodeT5}   & {\cellcolor[rgb]{0.875,0.875,0.875}}None & {\cellcolor[rgb]{0.875,0.875,0.875}}100   & {\cellcolor[rgb]{0.875,0.875,0.875}}100   & {\cellcolor[rgb]{0.875,0.875,0.875}}100   & {\cellcolor[rgb]{0.875,0.875,0.875}}100   & {\cellcolor[rgb]{0.875,0.875,0.875}}100 & {\cellcolor[rgb]{0.875,0.875,0.875}}99.76    & {\cellcolor[rgb]{0.875,0.875,0.875}}100 & {\cellcolor[rgb]{0.875,0.875,0.875}}93.07   & {\cellcolor[rgb]{0.875,0.875,0.875}}99.10  \\
                          & SS                                       & 66.69                                     & 52.19                                     & 65.02                                     & 75.62                                     & 76.25                                   & 81.59                                        & \uline{66.14}                           & \uline{71.79}                               & 69.41                                      \\
                          & AC                                       & 83.59                                     & 84.06                                     & 72.75                                     & 82.55                                     & 99.92                                   & 88.37                                        & 79.76                                   & 85.74                                       & 84.59                                      \\
                          & ONION                                    & \uline{54.74}                             & \uline{51.79}                             & \uline{64.22}                             & \uline{55.78}                             & 67.73                                   & \uline{52.59}                                & 93.39                                   & 73.47                                       & \uline{64.21}                              \\
                          & OSeqL                & 69.08                                     & {72.35}                 & 74.74                                     & {66.37}                 & \uline{66.45}                           & {65.02}                    & 95.94                                   & {88.61}                   & 74.82                                      \\
                          & {\tool}                                     & \textbf{2.95}                             & \textbf{4.54}                             & \textbf{2.23}                             & \textbf{5.26}                             & \textbf{1.27}                           & \textbf{5.34}                                & \textbf{1.51}                           & \textbf{11.47}                              & \textbf{4.32}                              \\ 
\hline\hline
\multicolumn{11}{c}{Clone Detection}                                                                                                                                                                                                                                                                                                                                                                                                                                               \\ 
\hline
\multirow{6}{*}{CodeBERT} & {\cellcolor[rgb]{0.875,0.875,0.875}}None & {\cellcolor[rgb]{0.875,0.875,0.875}}100   & {\cellcolor[rgb]{0.875,0.875,0.875}}100   & {\cellcolor[rgb]{0.875,0.875,0.875}}100   & {\cellcolor[rgb]{0.875,0.875,0.875}}100   & {\cellcolor[rgb]{0.875,0.875,0.875}}100 & {\cellcolor[rgb]{0.875,0.875,0.875}}100      & {\cellcolor[rgb]{0.875,0.875,0.875}}100 & {\cellcolor[rgb]{0.875,0.875,0.875}}100     & {\cellcolor[rgb]{0.875,0.875,0.875}}100    \\
                          & SS                                       & 100                                       & 100                                       & 100                                       & 100                                       & 100                                     & 100                                          & 100                                     & 100                                         & 100                                        \\
                          & AC                                       & 100                                       & 100                                       & 100                                       & 100                                       & 100                                     & 100                                          & \uline{27.92}                           & 100                                         & 90.99                                      \\
                          & ONION                                    & 99.7                                      & 99.62                                     & 96.91                                     & 97.07                                     & 89.77                                   & 90.52                                        & 99.85                                   & 99.85                                       & 96.66                                      \\
                          & OSeqL                & \uline{74.49}                             & {\uline{65.09}}         & \textbf{9.71}                             & {\uline{18.96}}         & \uline{53.57}                           & {\uline{39.95}}            & 96.99                                   & {\uline{99.47}}           & \uline{57.28}                              \\
                          & {\tool}                                     & \textbf{12.87}                            & \textbf{14.15}                            & \uline{11.59}                             & \textbf{17.68}                            & \textbf{22.87}                          & \textbf{23.55}                               & \textbf{21.22}                          & \textbf{19.94}                              & \textbf{17.98}                             \\ 
\hline
\multirow{6}{*}{CodeT5}   & {\cellcolor[rgb]{0.875,0.875,0.875}}None & {\cellcolor[rgb]{0.875,0.875,0.875}}100   & {\cellcolor[rgb]{0.875,0.875,0.875}}100   & {\cellcolor[rgb]{0.875,0.875,0.875}}100   & {\cellcolor[rgb]{0.875,0.875,0.875}}100   & {\cellcolor[rgb]{0.875,0.875,0.875}}100 & {\cellcolor[rgb]{0.875,0.875,0.875}}100      & {\cellcolor[rgb]{0.875,0.875,0.875}}100 & {\cellcolor[rgb]{0.875,0.875,0.875}}100     & {\cellcolor[rgb]{0.875,0.875,0.875}}100    \\
                          & SS                                       & 100                                       & 100                                       & 100                                       & 100                                       & 100                                     & 100                                          & 100                                     & 100                                         & 100                                        \\
                          & AC                                       & 100                                       & 100                                       & 24.08                                     & 100                                       & 100                                     & 100                                          & \uline{21.82}                           & 100                                         & 80.74                                      \\
                          & ONION                                    & 96.69                                     & 99.7                                      & 97.07                                     & 96.76                                     & 93.75                                   & 88.64                                        & 91.85                                   & 99.92                                       & 95.55                                      \\
                          & OSeqL                & \uline{76.00}                             & {\uline{75.02}}         & \textbf{17.46}                            & {\uline{47.93}}         & \uline{30.02}                           & {\textbf{14.00}}           & 98.95                                   & {\uline{99.32}}           & \uline{57.34}                              \\
                          & {\tool}                                     & \textbf{16.55}                            & \textbf{11.74}                            & \uline{19.86}                             & \textbf{16.33}                            & \textbf{19.86}                          & \uline{18.36}                                & \textbf{17.38}                          & \textbf{14.82}                              & \textbf{16.86}                             \\ 
\hline\hline
\multicolumn{11}{c}{Program Repair}                                                                                                                                                                                                                                                                                                                                                                                                                                                \\ 
\hline
\multirow{6}{*}{CodeBERT} & {\cellcolor[rgb]{0.875,0.875,0.875}}None & {\cellcolor[rgb]{0.875,0.875,0.875}}98.06 & {\cellcolor[rgb]{0.875,0.875,0.875}}100   & {\cellcolor[rgb]{0.875,0.875,0.875}}100   & {\cellcolor[rgb]{0.875,0.875,0.875}}98.17 & {\cellcolor[rgb]{0.875,0.875,0.875}}100 & {\cellcolor[rgb]{0.875,0.875,0.875}}100      & {\cellcolor[rgb]{0.875,0.875,0.875}}100 & {\cellcolor[rgb]{0.875,0.875,0.875}}100     & {\cellcolor[rgb]{0.875,0.875,0.875}}99.53  \\
                          & SS                                       & 97.83                                     & 98.29                                     & 97.83                                     & 98.06                                     & 97.94                                   & 98.17                                        & 4.91                                    & \uline{6.06}                                & 74.89                                      \\
                          & AC                                       & \uline{1.26}                              & \uline{1.37}                              & \textbf{1.26}                             & 97.94                                     & \uline{1.60}                            & \textbf{1.26}                                & \textbf{0}                              & 12.11                                       & \uline{14.60}                              \\
                          & ONION                                    & 50.06                                     & 52.57                                     & 41.94                                     & \uline{39.66}                             & 23.89                                   & \uline{24.57}                                & 74.40                                   & 73.83                                       & 47.62                                      \\
                          & OSeqL                & \textbackslash{}                          & {\textbackslash{}}      & \textbackslash{}                          & {\textbackslash{}}      & \textbackslash{}                        & {\textbackslash{}}         & \textbackslash{}                        & {\textbackslash{}}        & \textbackslash{}                           \\
                          & {\tool}                                     & \textbf{0.46}                             & \textbf{0.46}                             & \uline{3.20}                              & \textbf{2.97}                             & \textbf{0.69}                           & \textbf{1.26}                                & \uline{4.11}                            & \textbf{5.94}                               & \textbf{2.39}                              \\ 
\hline
\multirow{6}{*}{CodeT5}   & {\cellcolor[rgb]{0.875,0.875,0.875}}None & {\cellcolor[rgb]{0.875,0.875,0.875}}98.17 & {\cellcolor[rgb]{0.875,0.875,0.875}}97.71 & {\cellcolor[rgb]{0.875,0.875,0.875}}98.06 & {\cellcolor[rgb]{0.875,0.875,0.875}}97.83 & {\cellcolor[rgb]{0.875,0.875,0.875}}100 & {\cellcolor[rgb]{0.875,0.875,0.875}}100      & {\cellcolor[rgb]{0.875,0.875,0.875}}100 & {\cellcolor[rgb]{0.875,0.875,0.875}}100     & {\cellcolor[rgb]{0.875,0.875,0.875}}98.97  \\
                          & SS                                       & 97.71                                     & 97.71                                     & 97.37                                     & 97.83                                     & 97.03                                   & 96.00                                        & \textbf{0}                              & \textbf{0}                                  & 72.96                                      \\
                          & AC                                       & \uline{1.14}                              & \textbf{1.26}                             & \uline{1.37}                              & 97.37                                     & \textbf{0}                              & \textbf{1.37}                                & \textbf{0}                              & 14.90                                       & \uline{14.68}                              \\
                          & ONION                                    & 54.40                                     & 53.14                                     & 68.57                                     & \uline{66.17}                             & 47.31                                   & 39.09                                        & 74.51                                   & 74.40                                       & 59.70                                      \\
                          & OSeqL                & \textbackslash{}                          & {\textbackslash{}}      & \textbackslash{}                          & {\textbackslash{}}      & \textbackslash{}                        & {\textbackslash{}}         & \textbackslash{}                        & {\textbackslash{}}        & \textbackslash{}                           \\
                          & {\tool}                                     & \textbf{0.69}                             & \uline{1.40}                              & \textbf{1.14}                             & \textbf{3.09}                             & \uline{0.80}                            & \uline{2.29}                                 & \uline{5.83}                            & \uline{5.26}                                & \textbf{2.56}                              \\
\bottomrule
\end{tabular}
}
\end{table*}

\subsubsection{\textbf{Setup}}
\label{sec:attack_process}
We utilize four backdoor attacks, i.e., BNC (Fixed), BNC (Grammar), CodePoisoner (Dead-code), and CodePoisoner (identifier),
to attack CodeBERT and CodeT5 models across three tasks with 5\% and 1\% poisoning rates, respectively. Then, four defense baselines and {\tool} are leveraged to defend the attacks on the victim models. We report the Attack Success Rate (ASR) of the victim models without any defense and the ASR\textsubscript{D} of the models with the defense methods applied.

\subsubsection{\textbf{Results}}
Table \ref{tab:rq1} presents the defense performance of four baselines and {\tool} against four backdoor attacks. Overall, {\tool} effectively mitigates all the backdoor attacks, reducing the average ASR of the victim models across three tasks from a range of 98.97\%-100\% (no defense applied) to just 2.39\%-44.24\%. In six experimental groups, spanning two victim models and three tasks, {\tool} achieves an average ASR\textsubscript{D} of less than 5\% in three of these groups. Besides, {\tool} demonstrates robust performance across different poisoning rates (1\% and 5\%), proving effective even under varying levels of attack severity.

Among all the evaluated defense methods, {\tool} is the most effective, achieving the lowest ASR\textsubscript{D} in most attack scenarios. For defect detection and clone detection tasks, {\tool} consistently outperforms other baselines in defending against the four attack methods with different settings. Compared to the best-performing baselines, for the defect detection task, {\tool} substantially reduces the average ASR\textsubscript{D} by 19.60\% and 59.89\% on CodeBERT and CodeT5, respectively. For the clone detection task, {\tool} significantly reduces the average ASR\textsubscript{D} by 39.30\% on CodeBERT and 40.48\% on CodeT5.
Regarding program repair, although the ASR\textsubscript{D} of {\tool} is slightly higher than that of AC or SS in some attack scenarios, {the difference is relatively small, and } the average ASR\textsubscript{D} of {\tool} is substantially lower than both baselines. Compared to the best-performing baseline, {\tool} lowers the average ASR\textsubscript{D} by 12.21\% on CodeBERT and 12.12\% on CodeT5 {for program repair task.} 

We attribute these results to {\tool}'s ability to accurately identify and neutralize backdoor triggers within the poisoned code snippets. By employing identifier-level and statement-level masking strategies and calculating suspicion scores for all masked elements, {\tool} enhances its capacity to detect subtle triggers that might be overlooked by other defense methods.


\begin{table*}[htbp]
\vspace{-0.3cm}
\centering
\caption{The performance of the victim models on the clean samples when employing different defense methods. For defect detection and clone detection tasks, we report ACC\textsubscript{D} (\%); for program repair task, the performance of BLEU\textsubscript{D} (\%) is reported. \textit{None} indicates no defense applied. Higher ACC\textsubscript{D} or BLEU\textsubscript{D} indicate more effective defense methods. The best and second-best results are highlighted in bold and underlined.}
\label{tab:rq2}
\vspace{-0.1cm}
\resizebox{0.75\textwidth}{6cm}{
\begin{tabular}{c|c|cc|cc|cc|cc|c} 
\toprule
\multirow{2}{*}{Model}    & \multirow{2}{*}{Defense}                 & \multicolumn{2}{c|}{\begin{tabular}[c]{@{}c@{}}BNC\\(Fixed)\end{tabular}}             & \multicolumn{2}{c|}{\begin{tabular}[c]{@{}c@{}}BNC\\(Grammar)\end{tabular}}           & \multicolumn{2}{c|}{\begin{tabular}[c]{@{}c@{}}CodePoisoner\\(Dead-code)\end{tabular}} & \multicolumn{2}{c|}{\begin{tabular}[c]{@{}c@{}}CodePoisoner\\(Identifier)\end{tabular}} & \multirow{2}{*}{Average}                   \\
                          &                                          & $\alpha$=5\%                              & $\alpha$=1\%                              & $\alpha$=5\%                              & $\alpha$=1\%                              & $\alpha$=5\%                              & $\alpha$=1\%                               & $\alpha$=5\%                              & $\alpha$=1\%                              &                                            \\ 
\hline\hline
{}      & \multicolumn{10}{c}{Defect Detection}                                                                                                                                                                                                                                                                                                                                                                                                                  \\ 
\hline
\multirow{6}{*}{CodeBERT} & {\cellcolor[rgb]{0.875,0.875,0.875}}None & {\cellcolor[rgb]{0.875,0.875,0.875}}61.38 & {\cellcolor[rgb]{0.875,0.875,0.875}}62.15 & {\cellcolor[rgb]{0.875,0.875,0.875}}62.88 & {\cellcolor[rgb]{0.875,0.875,0.875}}63.65 & {\cellcolor[rgb]{0.875,0.875,0.875}}62.63 & {\cellcolor[rgb]{0.875,0.875,0.875}}62.66  & {\cellcolor[rgb]{0.875,0.875,0.875}}62.45 & {\cellcolor[rgb]{0.875,0.875,0.875}}62.33 & {\cellcolor[rgb]{0.875,0.875,0.875}}62.52  \\
                          & SS                                       & \textbf{62.88}                            & \textbf{63.51}                            & \uline{62.41}                             & \uline{62.77}                             & \uline{62.11}                             & \textbf{63.40}                             & \textbf{61.82}                            & \uline{62.19}                             & \textbf{62.64}                             \\
                          & AC                                       & 56.70                                     & 56.08                                     & 59.22                                     & 57.65                                     & 58.38                                     & 55.23                                      & 59.85                                     & 56.27                                     & 57.42                                      \\
                          & ONION                                    & 58.49                                     & 57.32                                     & 58.75                                     & 57.65                                     & 58.86                                     & 57.83                                      & 57.14                                     & 58.16                                     & 58.03                                      \\
                          & OSeqL                & 58.02                                     & {56.30}                 & 59.33                                     & {60.40}                 & 57.14                                     & {56.70}                  & 58.86                                     & {57.14}                 & 57.99                                      \\
                          &  {\tool}                                        & \uline{61.27}                             & \uline{62.15}                             & \textbf{62.88}                            & \textbf{62.96}                            & \textbf{62.37}                            & \uline{62.66}                              & \uline{61.68}                             & \textbf{62.33}                            & \uline{62.29}                              \\ 
\hline
\multirow{6}{*}{CodeT5}   & {\cellcolor[rgb]{0.875,0.875,0.875}}None & {\cellcolor[rgb]{0.875,0.875,0.875}}60.21 & {\cellcolor[rgb]{0.875,0.875,0.875}}60.47 & {\cellcolor[rgb]{0.875,0.875,0.875}}60.32 & {\cellcolor[rgb]{0.875,0.875,0.875}}58.67 & {\cellcolor[rgb]{0.875,0.875,0.875}}61.42 & {\cellcolor[rgb]{0.875,0.875,0.875}}62.19  & {\cellcolor[rgb]{0.875,0.875,0.875}}61.60 & {\cellcolor[rgb]{0.875,0.875,0.875}}59.44 & {\cellcolor[rgb]{0.875,0.875,0.875}}60.54  \\
                          & SS                                       & \uline{59.19}                             & \uline{59.48}                             & \uline{60.14}                             & \textbf{61.60}                            & \uline{59.30}                             & \textbf{61.24}                             & \uline{59.55}                             & \textbf{60.72}                            & \textbf{60.15}                             \\
                          & AC                                       & 56.48                                     & 56.88                                     & 56.63                                     & 56.88                                     & 50.04                                     & 56.59                                      & 58.57                                     & \uline{56.88}                             & 56.12                                      \\
                          & ONION                                    & 56.03                                     & 56.66                                     & 58.78                                     & 56.52                                     & 57.65                                     & 53.62                                      & 57.17                                     & 54.87                                     & 56.41                                      \\
                          & OSeqL                & 54.81                                     & {51.82}                 & 52.21                                     & {53.67}                 & 49.71                                     & {53.32}                  & 52.06                                     & {50.57}                 & 52.27                                      \\
                          &   {\tool}                                       & \textbf{60.18}                            & \textbf{59.77}                            & \textbf{60.25}                            & \uline{58.38}                             & \textbf{61.38}                            & \uline{58.71}                              & \textbf{61.53}                            & 56.30                                     & \uline{59.56}                              \\ 
\hline\hline
\multicolumn{11}{c}{Clone Detection}                                                                                                                                                                                                                                                                                                                                                                                                                                               \\ 
\hline
\multirow{6}{*}{CodeBERT} & {\cellcolor[rgb]{0.875,0.875,0.875}}None & {\cellcolor[rgb]{0.875,0.875,0.875}}97.09 & {\cellcolor[rgb]{0.875,0.875,0.875}}96.96 & {\cellcolor[rgb]{0.875,0.875,0.875}}97.31 & {\cellcolor[rgb]{0.875,0.875,0.875}}97.02 & {\cellcolor[rgb]{0.875,0.875,0.875}}96.98 & {\cellcolor[rgb]{0.875,0.875,0.875}}96.57  & {\cellcolor[rgb]{0.875,0.875,0.875}}96.73 & {\cellcolor[rgb]{0.875,0.875,0.875}}96.79 & {\cellcolor[rgb]{0.875,0.875,0.875}}96.93  \\
                          & SS                                       & 96.41                                     & \textbf{96.82}                            & \uline{96.41}                             & \textbf{97.09}                            & \uline{96.42}                             & \textbf{96.88}                             & \uline{96.15}                             & 95.95                                     & \uline{96.52}                              \\
                          & AC                                       & 92.04                                     & 92.51                                     & 94.86                                     & 94.41                                     & 94.81                                     & 94.89                                      & 94.79                                     & 88.01                                     & 93.29                                      \\
                          & ONION                                    & \uline{96.45}                             & 95.60                                     & 95.72                                     & 94.86                                     & 96.05                                     & \uline{96.53}                              & 95.25                                     & \uline{96.48}                             & 95.87                                      \\
                          & OSeqL                & 94.51                                     & {94.02}                 & 95.58                                     & {96.12}                 & 94.82                                     & {94.07}                  & 93.98                                     & {95.32}                 & 94.80                                      \\
                          &  {\tool}                                        & \textbf{97.02}                            & \uline{96.25}                             & \textbf{97.11}                            & \uline{96.59}                             & \textbf{96.90}                            & 95.93                                      & \textbf{96.64}                            & \textbf{96.54}                            & \textbf{96.62}                             \\ 
\hline
\multirow{6}{*}{CodeT5}   & {\cellcolor[rgb]{0.875,0.875,0.875}}None & {\cellcolor[rgb]{0.875,0.875,0.875}}97.37 & {\cellcolor[rgb]{0.875,0.875,0.875}}97.40 & {\cellcolor[rgb]{0.875,0.875,0.875}}97.16 & {\cellcolor[rgb]{0.875,0.875,0.875}}97.36 & {\cellcolor[rgb]{0.875,0.875,0.875}}97.47 & {\cellcolor[rgb]{0.875,0.875,0.875}}97.51  & {\cellcolor[rgb]{0.875,0.875,0.875}}97.16 & {\cellcolor[rgb]{0.875,0.875,0.875}}97.01 & {\cellcolor[rgb]{0.875,0.875,0.875}}97.31  \\
                          & SS                                       & \textbf{96.89}                            & \textbf{97.25}                            & \textbf{96.85}                            & \textbf{97.40}                            & \textbf{97.51}                            & \textbf{97.03}                             & \textbf{97.11}                            & \textbf{97.06}                            & \textbf{97.14}                             \\
                          & AC                                       & 88.71                                     & 92.71                                     & 95.41                                     & 93.71                                     & 95.18                                     & 94.60                                      & \uline{95.45}                             & 88.27                                     & 93.01                                      \\
                          & ONION                                    & \uline{96.33}                             & \textbf{97.25}                            & \uline{97.02}                             & 95.38                                     & \uline{96.13}                             & 96.28                                      & 95.00                                     & 95.20                                     & \uline{96.07}                              \\
                          & OSeqL                & 94.32                                     & {93.89}                 & 93.16                                     & {93.39}                 & 95.20                                     & {95.45}                  & 94.49                                     & {96.01}                 & 94.49                                      \\
                          & {\tool}                                         & 95.72                                     & \uline{95.85}                             & 95.09                                     & \uline{96.32}                             & 95.55                                     & \uline{96.70}                              & 95.12                                     & \uline{96.20}                             & 95.82                                      \\ 
\hline\hline
\multicolumn{11}{c}{Program Repair}                                                                                                                                                                                                                                                                                                                                                                                                                                                \\ 
\hline
\multirow{6}{*}{CodeBERT} & {\cellcolor[rgb]{0.875,0.875,0.875}}None & {\cellcolor[rgb]{0.875,0.875,0.875}}83.54 & {\cellcolor[rgb]{0.875,0.875,0.875}}83.57 & {\cellcolor[rgb]{0.875,0.875,0.875}}82.84 & {\cellcolor[rgb]{0.875,0.875,0.875}}83.44 & {\cellcolor[rgb]{0.875,0.875,0.875}}83.58 & {\cellcolor[rgb]{0.875,0.875,0.875}}82.89  & {\cellcolor[rgb]{0.875,0.875,0.875}}83.44 & {\cellcolor[rgb]{0.875,0.875,0.875}}83.02 & {\cellcolor[rgb]{0.875,0.875,0.875}}83.29  \\
                          & SS                                       & \uline{83.08}                             & \textbf{83.37}                            & \textbf{83.20}                            & 83.10                                     & \uline{83.40}                             & \textbf{83.52}                             & \uline{83.20}                             & \textbf{83.44}                            & \textbf{83.29}                             \\
                          & AC                                       & 82.37                                     & \uline{82.36}                             & \uline{82.94}                             & \uline{83.21}                             & 83.31                                     & \uline{83.09}                              & 83.09                                     & \uline{83.32}                             & 82.96                                      \\
                          & ONION                                    & 46.45                                     & 47.27                                     & 49.24                                     & 50.5                                      & 47.06                                     & 46.45                                      & 46.72                                     & 48.39                                     & 47.76                                      \\
                          & OSeqL                & \textbackslash{}                          & {\textbackslash{}}      & \textbackslash{}                          & {\textbackslash{}}      & \textbackslash{}                          & {\textbackslash{}}       & \textbackslash{}                          & {\textbackslash{}}      & \textbackslash{}                           \\
                          & {\tool}                                         & \textbf{83.36}                            & \textbf{83.37}                            & 82.78                                     & \textbf{83.31}                            & \textbf{83.48}                            & 82.67                                      & \textbf{83.28}                            & 82.90                                     & \uline{83.14}                              \\ 
\hline
\multirow{6}{*}{CodeT5}   & {\cellcolor[rgb]{0.875,0.875,0.875}}None & {\cellcolor[rgb]{0.875,0.875,0.875}}83.54 & {\cellcolor[rgb]{0.875,0.875,0.875}}83.64 & {\cellcolor[rgb]{0.875,0.875,0.875}}83.57 & {\cellcolor[rgb]{0.875,0.875,0.875}}83.62 & {\cellcolor[rgb]{0.875,0.875,0.875}}83.50 & {\cellcolor[rgb]{0.875,0.875,0.875}}83.61  & {\cellcolor[rgb]{0.875,0.875,0.875}}83.64 & {\cellcolor[rgb]{0.875,0.875,0.875}}83.58 & {\cellcolor[rgb]{0.875,0.875,0.875}}83.59  \\
                          & SS                                       & \textbf{83.58}                            & \textbf{83.61}                            & \textbf{83.71}                            & \textbf{83.69}                            & \textbf{83.47}                            & \textbf{83.63}                             & \textbf{83.61}                            & \textbf{83.48}                            & \textbf{83.60}                             \\
                          & AC                                       & \uline{83.55}                             & \uline{83.74}                             & \uline{83.67}                             & \uline{83.56}                             & \uline{83.45}                             & \uline{83.49}                              & \uline{83.63}                             & \uline{83.57}                             & \uline{83.58}                              \\
                          & ONION                                    & 57.43                                     & 57.56                                     & 57.85                                     & 57.12                                     & 57.23                                     & 56.82                                      & 58.14                                     & 57.28                                     & 57.43                                      \\
                          & OSeqL                & \textbackslash{}                          & {\textbackslash{}}      & \textbackslash{}                          & {\textbackslash{}}      & \textbackslash{}                          & {\textbackslash{}}       & \textbackslash{}                          & {\textbackslash{}}      & \textbackslash{}                           \\
                          &  {\tool}                                        & 83.43                                     & 83.50                                     & 83.51                                     & 83.48                                     & 83.35                                     & 83.44                                      & 83.42                                     & 83.46                                     & 83.45                                      \\
\bottomrule
\end{tabular}
\vspace{-0.6cm}
}
\end{table*}

\mybox{
\textbf{Answering RQ1:} {\tool} not only surpasses the baseline methods in defense effectiveness but also maintains high robustness across different attack scenarios and poisoning rates.
}

\subsection{RQ2: Impacts on the Functionalities of the Victim Models}

\subsubsection{\textbf{Setup}}
We use the same experimental settings as in RQ1, but with different evaluation metrics. For the defect detection and clone detection tasks, we use ACC\textsubscript{D} to evaluate the performance of {\tool}. For the program repair task, we use BLEU\textsubscript{D} as the evaluation metric.

\subsubsection{\textbf{Results}}
Table \ref{tab:rq2} illustrates the performance of victim models using different defense methods on clean samples.  Overall, all defense methods negatively impact the performance of the victim models on clean samples; however, the negative effects of {\tool} are minimal. Across six experimental groups involving two victim models and three tasks, the performance decreases caused by {\tool} are only 0.23\% and 0.98\% in ACC\textsubscript{D} on CodeBERT and CodeT5 for defect detection, 0.31\% and 1.49\% in ACC\textsubscript{D} for clone detection, and 0.15\% and 0.14\% in BLEU\textsubscript{D} for program repair.  In three of the six experimental groups, {\tool} achieves the best or second-best results, indicating that it minimally affects the semantics of clean samples.  Specifically, {\tool} leverages entropy-based analysis to distinguish between poisoned and clean code, effectively isolating suspicious elements with high precision.  This capability ensures that only the malicious triggers are targeted and neutralized, while the benign elements of the code remain unaffected, thereby preserving the overall functionality and correctness of the victim models.

\mybox{
\textbf{Answering RQ2:} {\tool} only has a slight impact on the performance of models on clean samples, indicating its effectiveness in preserving semantic integrity during the process of eliminating identified triggers from the inputs.} 


\subsection{RQ3: Ablation Study}
\label{RQ3}

\begin{figure*}[t]
\centering
\includegraphics[width=0.8\textwidth]{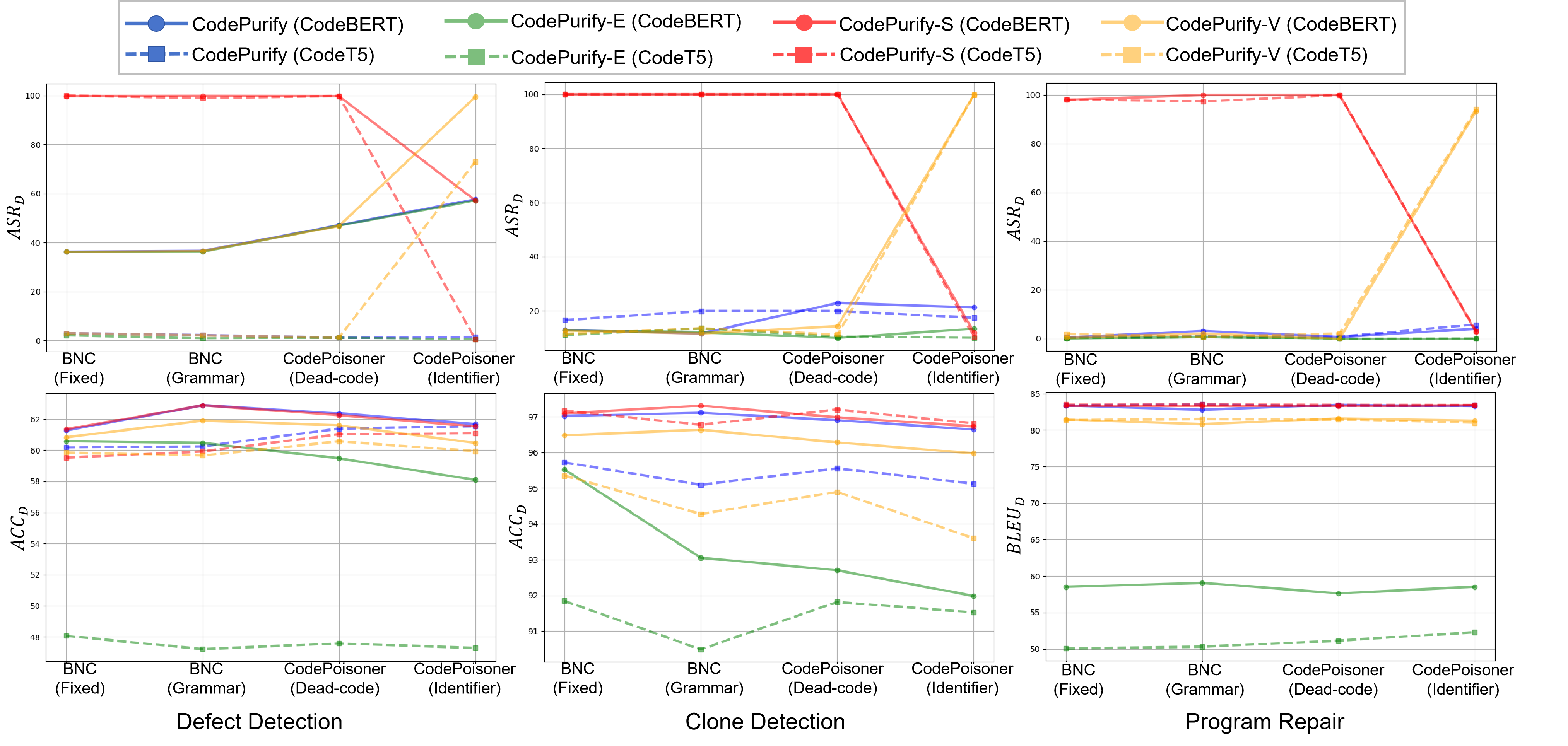}
\vspace{-0.1cm}
\caption{Ablation study on {\tool}. The top sub-figures show the ASR\textsubscript{D} of {\tool} and its variants, where lower ASR\textsubscript{D} indicates better defense. The bottom sub-figures depict the ACC\textsubscript{D} or BLEU\textsubscript{D} of {\tool} and its variants, where higher ACC\textsubscript{D} or BLEU\textsubscript{D} indicates better defense.
}
\vspace{-0.2cm}
\label{fig:ablation}
\end{figure*}

\subsubsection{\textbf{Setup}}
We conduct an ablation study using three variants of {\tool}: (1) \textit{{\tool}-E}, which omits entropy analysis and purifies all input code regardless of whether it is poisoned or clean; (2) \textit{{\tool}-S}, which excludes statement-level masking and focuses solely on identifier-level code purification; and (3) \textit{{\tool}-V}, which excludes identifier-level masking and focuses solely on statement-level code purification. For a fair comparison, {\tool} and three variants follow the same experimental setup.

\subsubsection{\textbf{Results}}
Figure \ref{fig:ablation} illustrates the performance of {\tool} and its three variants against four backdoor attacks with a 5\% poisoning rate across three tasks and two victim models. {Due to space limits, we do not present the performance with a 1\% poisoning rate considering it exerts a similar trend.}
The top sub-figures show the performance of ASR\textsubscript{D}, while the bottom sub-figures depict the performance of ACC\textsubscript{D} or BLEU\textsubscript{D}. Overall, {\tool} achieves the best balance between reducing attack success rates on poisoned samples and preserving the functionality of the victim models.

We observe that removing statement-level masking (\textit{{\tool}-S}) significantly reduces defensive effectiveness against dead-code insertion attacks. Similarly, without identifier-level masking (\textit{{\tool}-V}), the defense is nearly ineffective against identifier renaming attacks. Notably, \textit{{\tool}-S} shows higher ACC\textsubscript{D} and BLEU\textsubscript{D} than \textit{{\tool}-V}, as identifier renaming typically retains code semantics, while statement-level replacements often lead to semantic loss. We also find that excluding entropy analysis (\textit{{\tool}-E}) slightly improves defense performance, as it purifies all input code. However, this significantly degrades performance on clean samples. For instance, in the program repair task, \textit{{\tool}-E} reduces the average BLEU score by 25\% on CodeBERT and 32\% on CodeT5 compared to {\tool}, highlighting the importance of entropy analysis in maintaining clean sample performance.

Based on the above findings, we provide the following suggestions for selecting a suitable variant of {\tool} for specific attack scenarios:
(1) {\tool} is the best choice when the type of incoming backdoor attacks is unknown. It effectively defends against both dead-code insertion and identifier renaming attacks, which are the dominant forms of attack in the existing literature.
(2) When facing specific backdoor attacks, selecting a specialized variant of {\tool} (e.g., \textit{{\tool}-S} for statement-level attacks or \textit{{\tool}-V} for identifier-level attacks) may not significantly enhance defense performance but can better preserve the performance on clean samples.

\mybox{
\textbf{Answering RQ3:} Compared with its variants, {\tool} achieves the best balance between reducing attack success rates on poisoned samples and preserving the overall functionality of the victim models.}  
\section{Discussion}
\label{sec:discussion}


\subsection{Effect of Entropy Threshold}
\begin{figure}[t]
\centering
\includegraphics[width=0.8\columnwidth]{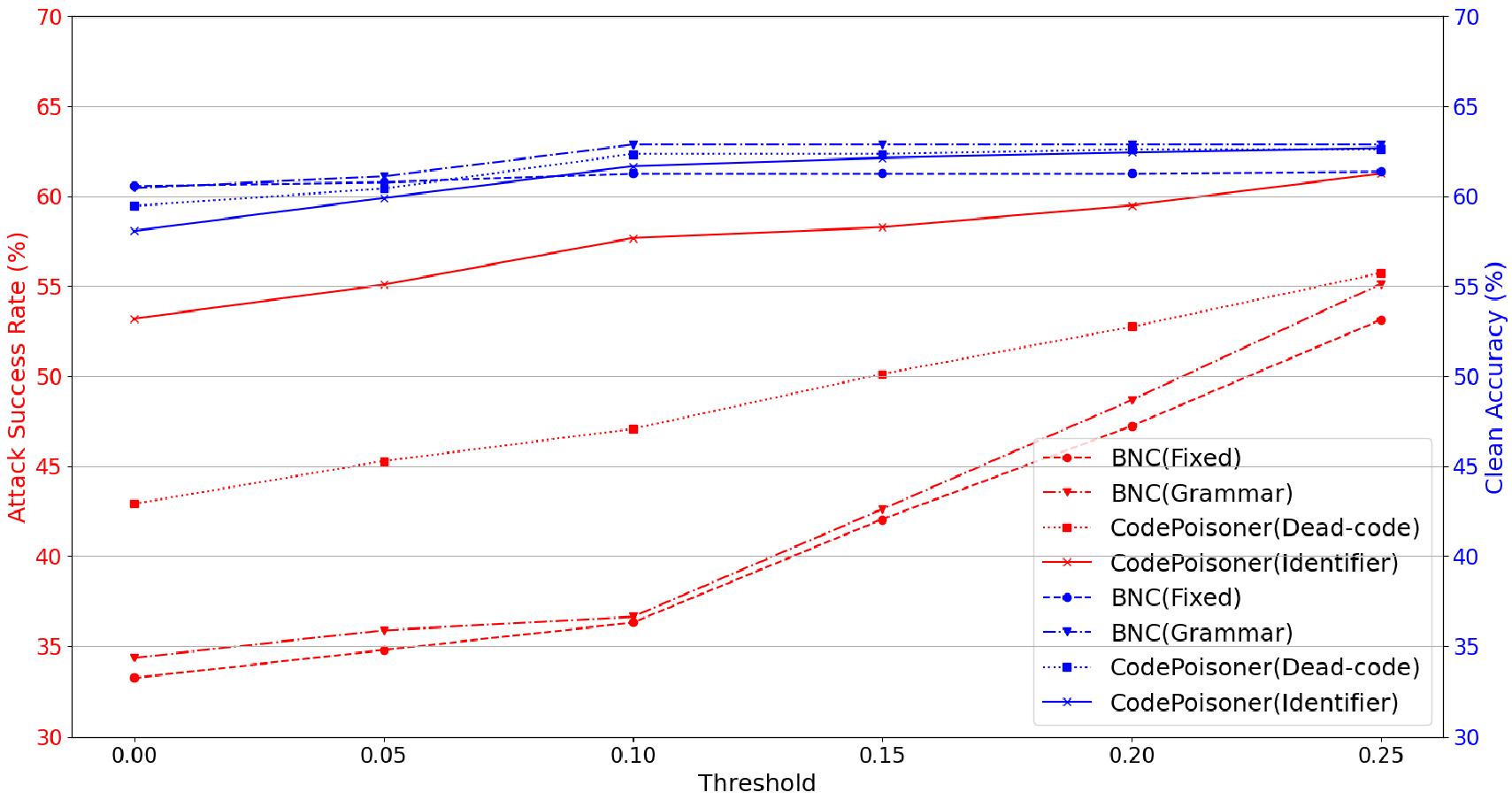}
\caption{Attack success rate (ASR\textsubscript{D}) and clean accuracy (ACC\textsubscript{D}) of {\tool} on defect detection task with different entropy thresholds ($t$).}
\vspace{-0.3cm}
\label{fig:threshold}
\end{figure}

The entropy threshold $t$ is the sole hyperparameter of {\tool}. In this section, we examine its impact on the defense performance. Figure \ref{fig:threshold} presents the ASR\textsubscript{D} and ACC\textsubscript{D} of {\tool} {when defensing four attack techniques} in the defect detection task with various values of $t$.
{Note that, we do not present the performance of other two tasks due to space limit and similar trends.}

The results demonstrate that a threshold of $t=0.1$ achieves the optimal balance between ASR\textsubscript{D} and ACC\textsubscript{D}, where a lower ASR\textsubscript{D} is desirable, and a higher ACC\textsubscript{D} is better. Specifically, when $t$ increases from 0 to 0.1, there is a significant rise in ACC\textsubscript{D}, while further increases in $t$ result in a much slower growth in ACC\textsubscript{D}. Therefore, if a clean validation set is unavailable, we recommend setting the threshold to 0.1.


\subsection{Time Efficiency}

We further evaluate the time efficiency of {\tool} and the baseline defenses. Table \ref{tab:time} presents the average time costs for defending against various backdoor attacks. Since SS and AC are training-time defenses, we additionally record their time costs during the {training} process, which involves detecting poisoned samples in the training sets and re-training the victim models. 

From the table, we can see that SS and AC, being training-time defenses, incur additional time costs for detecting poisoned samples and re-training victim models. This re-training process can be substantial when model parameters and training data are large. In contrast, inference-time defenses like ONION, OSeqL, and {\tool} add time overhead during the inference stage. However, {\tool} demonstrates superior time efficiency compared to ONION and OSeqL. For instance, in defect detection, {\tool} is 11 $\times$ faster than ONION and 2 $\times$ faster than OSeqL. {\tool}'s additional time cost mainly arises from its two-level masking strategies and subsequent calculation of suspicion scores. However, this overhead can be partially mitigated through batch processing, enhancing its overall efficiency.


\begin{table}
\centering
\caption{Average time spent defending against backdoor attacks. \textit{Training} denotes the time spent on training-time defense, which involves detecting poisoned samples from the training sets and re-training the models. \textit{Inference} denotes the time spent defending a test sample at inference time.}
\label{tab:time}
\vspace{-0.1cm}
\resizebox{0.9\columnwidth}{!}{
\begin{tabular}{ccccccc} 
\toprule
\multirow{2}{*}{Defense} & \multicolumn{2}{c}{Defect Detection} & \multicolumn{2}{c}{Clone Detection} & \multicolumn{2}{c}{Program Repair}  \\
                         & Training         & Inference         & Training         & Inference        & Training         & Inference        \\ 
\hline
None                     & \textbackslash{} & 0.01s             & \textbackslash{} & 0.02s            & \textbackslash{} & 0.34s            \\
SS                       & 1h               & 0.01s             & 3h               & 0.02s            & 24h              & 0.34s            \\
AC                       & 1h               & 0.01s             & 3h               & 0.02s            & 24h              & 0.34s            \\
ONION                    & \textbackslash{} & 2.47s             & \textbackslash{} & 1.86s            & \textbackslash{} & 1.98s            \\
OSeqL                    & \textbackslash{} & 0.45s             & \textbackslash{} & 1.07s            & \textbackslash{} & \textbackslash{}           \\
{\tool}                     & \textbackslash{} & 0.21s             & \textbackslash{} & 0.23s            & \textbackslash{} & 0.82s            \\
\bottomrule
\end{tabular}
\vspace{-0.3cm}
}
\end{table}

\subsection{Potential Applications}
The direct application of {\tool} is as a test-time backdoor defense. In this scenario, the victim model may already be compromised by a backdoor, and our defense aims to prevent malicious inputs from triggering it. When a user inputs a code snippet, {\tool} first determines if the code is poisoned. If it is, {\tool} locates the trigger and generates purified code by replacing the trigger with a benign code element. The purified code is then fed into the victim model. If the input is deemed clean, it is directly fed into the victim model.

Another potential application of {\tool} is as a training-time defense. 
As mentioned above, there are two categories of defense strategies, i.e., training-time defenses such as SS \cite{DBLP:conf/nips/Tran0M18} and AC \cite{chen2018detecting}, and test-time defenses such as ONION \cite{DBLP:conf/emnlp/QiCLYLS21} and OSeqL \cite{DBLP:journals/corr/abs-2312-04004}.
For training-time defense, the methods have access to the poisoned training dataset and aim to remove the poisoned samples.
Although our proposed {\tool} is evaluated in a test-time defense scenario, we think it can also be utilized for training-time defense considering its proficiency in detecting poisoned samples.
Specifically, for each code in the poisoned training dataset, {\tool} can employ masking strategies to generate multiple masked versions. It then computes suspicion scores and determines the presence of poisoning based on the entropy of these scores. If a code snippet is identified as poisoned, {\tool} eliminates it from the training set. The sanitized training dataset is then used to train the code model. 
We defer the experimental validation to future work.

\section{Related Work}
\label{sec:related}
Backdoor attacks, also known as trojan attacks \cite{DBLP:conf/ndss/LiuMALZW018}, have been extensively studied in the fields of computer vision (CV) \cite{DBLP:journals/corr/abs-1708-06733, DBLP:journals/corr/abs-1712-05526, DBLP:conf/codaspy/ZhongLSZ020, DBLP:conf/aaai/SahaSP20, DBLP:conf/nips/DoanLL21} and natural language processing (NLP) \cite{DBLP:conf/acl/KuritaMN20, DBLP:conf/ccs/LiLDZXZL21, DBLP:conf/acl/QiLCZ0WS20, DBLP:conf/emnlp/QiCZLLS21, DBLP:conf/naacl/GanL0LM00GF22}. With the neural code models playing an increasingly important role in the field of software engineering, backdoor attacks and defense against the code model have received more and more attention. Most of the existing backdoor attacks on code models inject triggers into normal samples via dead-code insertion or identifier renaming. For example, Ramakrishnan {et al.} \cite{DBLP:conf/icpr/RamakrishnanA22} explore backdoor attacks for the neural models of source code and propose to utilize pieces of dead-code as backdoor triggers.
Li {et al}. \cite{li2023poison} propose CodePoisoner, a backdoor attack method consisting of three rule-based poisoning strategies and one language-model-guided strategy. Yang {et al.} \cite{yang2024stealthy} present AFRAIDOOR, a stealthy backdoor attack method that uses adversarial perturbations to obtain adaptive triggers, that is, the trigger is varied from sample to sample.

Compared to backdoor attacks on code models, defense is still in its early stages. Previous studies \cite{DBLP:conf/icpr/RamakrishnanA22, yang2024stealthy, DBLP:conf/acl/SunCTF0Z023, wan2022you} usually use the defense techniques proposed by other domains, e.g., spectral signature \cite{DBLP:conf/nips/Tran0M18}, activation clustering, and outlier word detection. 
To the best of our knowledge, there are only two backdoor defense methods designed specifically for code models, namely CodeDetector \cite{li2023poison} and OSeqL \cite{DBLP:journals/corr/abs-2312-04004}. 
Different from the existing research, {\tool} can precisely detect both identifier-level and statement-level triggers by employing two-level masking strategies and an entropy-based measure.
\section{Conclusion}

Attack and defense are akin to twin brothers, each enhancing the capabilities of the other through their ongoing struggle.
For code models, numerous backdoor attack techniques have been developed. However, due to the stealthy nature of these attacks, there are almost no viable defense techniques, posing a significant threat to the security of code models.
In this paper, we propose a defense method against backdoor attacks on code models, through source code purification.
We conduct extensive experiments on four state-of-the-art backdoor attack methods with two poisoning rates to attack two victim models across three representative software engineering tasks. The results demonstrate {\tool}'s superior performance over four state-of-the-art baselines.

\bibliographystyle{IEEEtran}
\balance
\bibliography{ref}

\begin{thebibliography}{10}
\providecommand{\url}[1]{#1}
\csname url@samestyle\endcsname
\providecommand{\newblock}{\relax}
\providecommand{\bibinfo}[2]{#2}
\providecommand{\BIBentrySTDinterwordspacing}{\spaceskip=0pt\relax}
\providecommand{\BIBentryALTinterwordstretchfactor}{4}
\providecommand{\BIBentryALTinterwordspacing}{\spaceskip=\fontdimen2\font plus
\BIBentryALTinterwordstretchfactor\fontdimen3\font minus \fontdimen4\font\relax}
\providecommand{\BIBforeignlanguage}[2]{{%
\expandafter\ifx\csname l@#1\endcsname\relax
\typeout{** WARNING: IEEEtran.bst: No hyphenation pattern has been}%
\typeout{** loaded for the language `#1'. Using the pattern for}%
\typeout{** the default language instead.}%
\else
\language=\csname l@#1\endcsname
\fi
#2}}
\providecommand{\BIBdecl}{\relax}
\BIBdecl

\bibitem{zhou2019devign}
Y.~Zhou, S.~Liu, J.~Siow, X.~Du, and Y.~Liu, ``Devign: Effective vulnerability identification by learning comprehensive program semantics via graph neural networks,'' \emph{Advances in neural information processing systems}, vol.~32, 2019.

\bibitem{azeem2019machine}
M.~I. Azeem, F.~Palomba, L.~Shi, and Q.~Wang, ``Machine learning techniques for code smell detection: A systematic literature review and meta-analysis,'' \emph{Information and Software Technology}, vol. 108, pp. 115--138, 2019.

\bibitem{jin2023inferfix}
M.~Jin, S.~Shahriar, M.~Tufano, X.~Shi, S.~Lu, N.~Sundaresan, and A.~Svyatkovskiy, ``Inferfix: End-to-end program repair with llms,'' in \emph{Proceedings of the 31st ACM Joint European Software Engineering Conference and Symposium on the Foundations of Software Engineering}, 2023, pp. 1646--1656.

\bibitem{mashhadi2021applying}
E.~Mashhadi and H.~Hemmati, ``Applying codebert for automated program repair of java simple bugs,'' in \emph{2021 IEEE/ACM 18th International Conference on Mining Software Repositories (MSR)}.\hskip 1em plus 0.5em minus 0.4em\relax IEEE, 2021, pp. 505--509.

\bibitem{shin2023good}
J.~Shin, M.~Wei, J.~Wang, L.~Shi, and S.~Wang, ``The good, the bad, and the missing: Neural code generation for machine learning tasks,'' \emph{ACM Transactions on Software Engineering and Methodology}, vol.~33, no.~2, pp. 1--24, 2023.

\bibitem{le2020deep}
T.~H. Le, H.~Chen, and M.~A. Babar, ``Deep learning for source code modeling and generation: Models, applications, and challenges,'' \emph{ACM Computing Surveys (CSUR)}, vol.~53, no.~3, pp. 1--38, 2020.

\bibitem{hassan2024rethinking}
A.~E. Hassan, G.~A. Oliva, D.~Lin, B.~Chen, Z.~Ming \emph{et~al.}, ``Rethinking software engineering in the foundation model era: From task-driven ai copilots to goal-driven ai pair programmers,'' \emph{arXiv preprint arXiv:2404.10225}, 2024.

\bibitem{yang2024robustness}
Z.~Yang, Z.~Sun, T.~Z. Yue, P.~Devanbu, and D.~Lo, ``Robustness, security, privacy, explainability, efficiency, and usability of large language models for code,'' \emph{arXiv preprint arXiv:2403.07506}, 2024.

\bibitem{DBLP:conf/icpr/RamakrishnanA22}
\BIBentryALTinterwordspacing
G.~Ramakrishnan and A.~Albarghouthi, ``Backdoors in neural models of source code,'' in \emph{26th International Conference on Pattern Recognition, {ICPR} 2022, Montreal, QC, Canada, August 21-25, 2022}.\hskip 1em plus 0.5em minus 0.4em\relax {IEEE}, 2022, pp. 2892--2899. [Online]. Available: \url{https://doi.org/10.1109/ICPR56361.2022.9956690}
\BIBentrySTDinterwordspacing

\bibitem{wan2022you}
Y.~Wan, S.~Zhang, H.~Zhang, Y.~Sui, G.~Xu, D.~Yao, H.~Jin, and L.~Sun, ``You see what i want you to see: poisoning vulnerabilities in neural code search,'' in \emph{Proceedings of the 30th ACM Joint European Software Engineering Conference and Symposium on the Foundations of Software Engineering}, 2022, pp. 1233--1245.

\bibitem{yang2024stealthy}
Z.~Yang, B.~Xu, J.~M. Zhang, H.~J. Kang, J.~Shi, J.~He, and D.~Lo, ``Stealthy backdoor attack for code models,'' \emph{IEEE Transactions on Software Engineering}, 2024.

\bibitem{li2023poison}
J.~Li, Z.~Li, H.~Zhang, G.~Li, Z.~Jin, X.~Hu, and X.~Xia, ``Poison attack and poison detection on deep source code processing models,'' \emph{ACM Transactions on Software Engineering and Methodology}, 2023.

\bibitem{DBLP:journals/corr/abs-2312-04004}
\BIBentryALTinterwordspacing
A.~Hussain, M.~R.~I. Rabin, T.~Ahmed, M.~A. Alipour, and B.~Xu, ``Occlusion-based detection of trojan-triggering inputs in large language models of code,'' \emph{CoRR}, vol. abs/2312.04004, 2023. [Online]. Available: \url{https://doi.org/10.48550/arXiv.2312.04004}
\BIBentrySTDinterwordspacing

\bibitem{DBLP:conf/sp/WangYSLVZZ19}
\BIBentryALTinterwordspacing
B.~Wang, Y.~Yao, S.~Shan, H.~Li, B.~Viswanath, H.~Zheng, and B.~Y. Zhao, ``Neural cleanse: Identifying and mitigating backdoor attacks in neural networks,'' in \emph{2019 {IEEE} Symposium on Security and Privacy, {SP} 2019, San Francisco, CA, USA, May 19-23, 2019}.\hskip 1em plus 0.5em minus 0.4em\relax {IEEE}, 2019, pp. 707--723. [Online]. Available: \url{https://doi.org/10.1109/SP.2019.00031}
\BIBentrySTDinterwordspacing

\bibitem{DBLP:journals/corr/abs-2303-15564}
\BIBentryALTinterwordspacing
T.~Sun, L.~Pang, C.~Chen, and H.~Ling, ``Mask and restore: Blind backdoor defense at test time with masked autoencoder,'' \emph{CoRR}, vol. abs/2303.15564, 2023. [Online]. Available: \url{https://doi.org/10.48550/arXiv.2303.15564}
\BIBentrySTDinterwordspacing

\bibitem{DBLP:conf/acsac/DoanAR20}
\BIBentryALTinterwordspacing
B.~G. Doan, E.~Abbasnejad, and D.~C. Ranasinghe, ``Februus: Input purification defense against trojan attacks on deep neural network systems,'' in \emph{{ACSAC} '20: Annual Computer Security Applications Conference, Virtual Event / Austin, TX, USA, 7-11 December, 2020}.\hskip 1em plus 0.5em minus 0.4em\relax {ACM}, 2020, pp. 897--912. [Online]. Available: \url{https://doi.org/10.1145/3427228.3427264}
\BIBentrySTDinterwordspacing

\bibitem{DBLP:conf/icse/MaleticM01}
\BIBentryALTinterwordspacing
J.~I. Maletic and A.~Marcus, ``Supporting program comprehension using semantic and structural information,'' in \emph{Proceedings of the 23rd International Conference on Software Engineering, {ICSE} 2001, 12-19 May 2001, Toronto, Ontario, Canada}, H.~A. M{\"{u}}ller, M.~J. Harrold, and W.~Sch{\"{a}}fer, Eds.\hskip 1em plus 0.5em minus 0.4em\relax {IEEE} Computer Society, 2001, pp. 103--112. [Online]. Available: \url{https://doi.org/10.1109/ICSE.2001.919085}
\BIBentrySTDinterwordspacing

\bibitem{DBLP:conf/emnlp/QiCLYLS21}
\BIBentryALTinterwordspacing
F.~Qi, Y.~Chen, M.~Li, Y.~Yao, Z.~Liu, and M.~Sun, ``{ONION:} {A} simple and effective defense against textual backdoor attacks,'' in \emph{Proceedings of the 2021 Conference on Empirical Methods in Natural Language Processing, {EMNLP} 2021, Virtual Event / Punta Cana, Dominican Republic, 7-11 November, 2021}.\hskip 1em plus 0.5em minus 0.4em\relax Association for Computational Linguistics, 2021, pp. 9558--9566. [Online]. Available: \url{https://doi.org/10.18653/v1/2021.emnlp-main.752}
\BIBentrySTDinterwordspacing

\bibitem{DBLP:conf/ccs/LiuLTMAZ19}
\BIBentryALTinterwordspacing
Y.~Liu, W.~Lee, G.~Tao, S.~Ma, Y.~Aafer, and X.~Zhang, ``{ABS:} scanning neural networks for back-doors by artificial brain stimulation,'' in \emph{Proceedings of the 2019 {ACM} {SIGSAC} Conference on Computer and Communications Security, {CCS} 2019, London, UK, November 11-15, 2019}, L.~Cavallaro, J.~Kinder, X.~Wang, and J.~Katz, Eds.\hskip 1em plus 0.5em minus 0.4em\relax {ACM}, 2019, pp. 1265--1282. [Online]. Available: \url{https://doi.org/10.1145/3319535.3363216}
\BIBentrySTDinterwordspacing

\bibitem{renyi1961measures}
A.~R{\'e}nyi, ``On measures of entropy and information,'' in \emph{Proceedings of the fourth Berkeley symposium on mathematical statistics and probability, volume 1: contributions to the theory of statistics}, vol.~4.\hskip 1em plus 0.5em minus 0.4em\relax University of California Press, 1961, pp. 547--562.

\bibitem{tree-sitter}
``tree-sitter,'' \url{https://tree-sitter.github.io/tree-sitter/}, 2024.

\bibitem{DBLP:conf/acl/SennrichHB16a}
\BIBentryALTinterwordspacing
R.~Sennrich, B.~Haddow, and A.~Birch, ``Neural machine translation of rare words with subword units,'' in \emph{Proceedings of the 54th Annual Meeting of the Association for Computational Linguistics, {ACL} 2016, August 7-12, 2016, Berlin, Germany, Volume 1: Long Papers}.\hskip 1em plus 0.5em minus 0.4em\relax The Association for Computer Linguistics, 2016. [Online]. Available: \url{https://doi.org/10.18653/v1/p16-1162}
\BIBentrySTDinterwordspacing

\bibitem{DBLP:conf/acl/NaC023}
\BIBentryALTinterwordspacing
C.~Na, Y.~Choi, and J.~Lee, ``{DIP:} dead code insertion based black-box attack for programming language model,'' in \emph{Proceedings of the 61st Annual Meeting of the Association for Computational Linguistics (Volume 1: Long Papers), {ACL} 2023, Toronto, Canada, July 9-14, 2023}, A.~Rogers, J.~L. Boyd{-}Graber, and N.~Okazaki, Eds.\hskip 1em plus 0.5em minus 0.4em\relax Association for Computational Linguistics, 2023, pp. 7777--7791. [Online]. Available: \url{https://doi.org/10.18653/v1/2023.acl-long.430}
\BIBentrySTDinterwordspacing

\bibitem{DBLP:journals/jei/BishopN07}
\BIBentryALTinterwordspacing
C.~M. Bishop and N.~M. Nasrabadi, ``\emph{Pattern Recognition and Machine Learning},'' \emph{J. Electronic Imaging}, vol.~16, no.~4, p. 049901, 2007. [Online]. Available: \url{https://doi.org/10.1117/1.2819119}
\BIBentrySTDinterwordspacing

\bibitem{feng2020codebert}
Z.~Feng, D.~Guo, D.~Tang, N.~Duan, X.~Feng, M.~Gong, L.~Shou, B.~Qin, T.~Liu, D.~Jiang \emph{et~al.}, ``Codebert: A pre-trained model for programming and natural languages,'' \emph{arXiv preprint arXiv:2002.08155}, 2020.

\bibitem{DBLP:conf/iclr/FriedAL0WSZYZL23}
\BIBentryALTinterwordspacing
D.~Fried, A.~Aghajanyan, J.~Lin, S.~Wang, E.~Wallace, F.~Shi, R.~Zhong, S.~Yih, L.~Zettlemoyer, and M.~Lewis, ``Incoder: {A} generative model for code infilling and synthesis,'' in \emph{The Eleventh International Conference on Learning Representations, {ICLR} 2023, Kigali, Rwanda, May 1-5, 2023}.\hskip 1em plus 0.5em minus 0.4em\relax OpenReview.net, 2023. [Online]. Available: \url{https://openreview.net/pdf?id=hQwb-lbM6EL}
\BIBentrySTDinterwordspacing

\bibitem{chatgpt}
OpenAI, ``{ChatGPT},'' \url{https://openai.com/blog/chatgpt/}, 2022.

\bibitem{yang2022natural}
Z.~Yang, J.~Shi, J.~He, and D.~Lo, ``Natural attack for pre-trained models of code,'' in \emph{Proceedings of the 44th International Conference on Software Engineering}, 2022, pp. 1482--1493.

\bibitem{DBLP:conf/sigsoft/XiaZ22}
\BIBentryALTinterwordspacing
C.~S. Xia and L.~Zhang, ``Less training, more repairing please: revisiting automated program repair via zero-shot learning,'' in \emph{Proceedings of the 30th {ACM} Joint European Software Engineering Conference and Symposium on the Foundations of Software Engineering, {ESEC/FSE} 2022, Singapore, Singapore, November 14-18, 2022}, A.~Roychoudhury, C.~Cadar, and M.~Kim, Eds.\hskip 1em plus 0.5em minus 0.4em\relax {ACM}, 2022, pp. 959--971. [Online]. Available: \url{https://doi.org/10.1145/3540250.3549101}
\BIBentrySTDinterwordspacing

\bibitem{DBLP:conf/naacl/DevlinCLT19}
\BIBentryALTinterwordspacing
J.~Devlin, M.~Chang, K.~Lee, and K.~Toutanova, ``{BERT:} pre-training of deep bidirectional transformers for language understanding,'' in \emph{Proceedings of the 2019 Conference of the North American Chapter of the Association for Computational Linguistics: Human Language Technologies, {NAACL-HLT} 2019, Minneapolis, MN, USA, June 2-7, 2019, Volume 1 (Long and Short Papers)}.\hskip 1em plus 0.5em minus 0.4em\relax Association for Computational Linguistics, 2019, pp. 4171--4186. [Online]. Available: \url{https://doi.org/10.18653/v1/n19-1423}
\BIBentrySTDinterwordspacing

\bibitem{DBLP:conf/icsm/SvajlenkoIKRM14}
\BIBentryALTinterwordspacing
J.~Svajlenko, J.~F. Islam, I.~Keivanloo, C.~K. Roy, and M.~M. Mia, ``Towards a big data curated benchmark of inter-project code clones,'' in \emph{30th {IEEE} International Conference on Software Maintenance and Evolution, Victoria, BC, Canada, September 29 - October 3, 2014}.\hskip 1em plus 0.5em minus 0.4em\relax {IEEE} Computer Society, 2014, pp. 476--480. [Online]. Available: \url{https://doi.org/10.1109/ICSME.2014.77}
\BIBentrySTDinterwordspacing

\bibitem{DBLP:conf/nips/LuGRHSBCDJTLZSZ21}
\BIBentryALTinterwordspacing
S.~Lu, D.~Guo, S.~Ren, J.~Huang, A.~Svyatkovskiy, A.~Blanco, C.~B. Clement, D.~Drain, D.~Jiang, D.~Tang, G.~Li, L.~Zhou, L.~Shou, L.~Zhou, M.~Tufano, M.~Gong, M.~Zhou, N.~Duan, N.~Sundaresan, S.~K. Deng, S.~Fu, and S.~Liu, ``Codexglue: {A} machine learning benchmark dataset for code understanding and generation,'' in \emph{Proceedings of the Neural Information Processing Systems Track on Datasets and Benchmarks 1, NeurIPS Datasets and Benchmarks 2021, December 2021, virtual}, J.~Vanschoren and S.~Yeung, Eds., 2021. [Online]. Available: \url{https://datasets-benchmarks-proceedings.neurips.cc/paper/2021/hash/c16a5320fa475530d9583c34fd356ef5-Abstract-round1.html}
\BIBentrySTDinterwordspacing

\bibitem{tufano2019empirical}
M.~Tufano, C.~Watson, G.~Bavota, M.~D. Penta, M.~White, and D.~Poshyvanyk, ``An empirical study on learning bug-fixing patches in the wild via neural machine translation,'' \emph{ACM Transactions on Software Engineering and Methodology (TOSEM)}, vol.~28, no.~4, pp. 1--29, 2019.

\bibitem{wang2021codet5}
Y.~Wang, W.~Wang, S.~Joty, and S.~C. Hoi, ``Codet5: Identifier-aware unified pre-trained encoder-decoder models for code understanding and generation,'' \emph{arXiv preprint arXiv:2109.00859}, 2021.

\bibitem{DBLP:conf/nips/Tran0M18}
\BIBentryALTinterwordspacing
B.~Tran, J.~Li, and A.~Madry, ``Spectral signatures in backdoor attacks,'' in \emph{Advances in Neural Information Processing Systems 31: Annual Conference on Neural Information Processing Systems 2018, NeurIPS 2018, December 3-8, 2018, Montr{\'{e}}al, Canada}, S.~Bengio, H.~M. Wallach, H.~Larochelle, K.~Grauman, N.~Cesa{-}Bianchi, and R.~Garnett, Eds., 2018, pp. 8011--8021. [Online]. Available: \url{https://proceedings.neurips.cc/paper/2018/hash/280cf18baf4311c92aa5a042336587d3-Abstract.html}
\BIBentrySTDinterwordspacing

\bibitem{chen2018detecting}
B.~Chen, W.~Carvalho, N.~Baracaldo, H.~Ludwig, B.~Edwards, T.~Lee, I.~Molloy, and B.~Srivastava, ``Detecting backdoor attacks on deep neural networks by activation clustering,'' \emph{arXiv preprint arXiv:1811.03728}, 2018.

\bibitem{radford2019language}
A.~Radford, J.~Wu, R.~Child, D.~Luan, D.~Amodei, I.~Sutskever \emph{et~al.}, ``Language models are unsupervised multitask learners,'' \emph{OpenAI blog}, vol.~1, no.~8, p.~9, 2019.

\bibitem{DBLP:conf/acl/LiLCX0023}
\BIBentryALTinterwordspacing
Y.~Li, S.~Liu, K.~Chen, X.~Xie, T.~Zhang, and Y.~Liu, ``Multi-target backdoor attacks for code pre-trained models,'' in \emph{Proceedings of the 61st Annual Meeting of the Association for Computational Linguistics (Volume 1: Long Papers), {ACL} 2023, Toronto, Canada, July 9-14, 2023}, A.~Rogers, J.~L. Boyd{-}Graber, and N.~Okazaki, Eds.\hskip 1em plus 0.5em minus 0.4em\relax Association for Computational Linguistics, 2023, pp. 7236--7254. [Online]. Available: \url{https://doi.org/10.18653/v1/2023.acl-long.399}
\BIBentrySTDinterwordspacing

\bibitem{DBLP:conf/ndss/LiuMALZW018}
\BIBentryALTinterwordspacing
Y.~Liu, S.~Ma, Y.~Aafer, W.~Lee, J.~Zhai, W.~Wang, and X.~Zhang, ``Trojaning attack on neural networks,'' in \emph{25th Annual Network and Distributed System Security Symposium, {NDSS} 2018, San Diego, California, USA, February 18-21, 2018}.\hskip 1em plus 0.5em minus 0.4em\relax The Internet Society, 2018. [Online]. Available: \url{https://www.ndss-symposium.org/wp-content/uploads/2018/02/ndss2018\_03A-5\_Liu\_paper.pdf}
\BIBentrySTDinterwordspacing

\bibitem{DBLP:journals/corr/abs-1708-06733}
\BIBentryALTinterwordspacing
T.~Gu, B.~Dolan{-}Gavitt, and S.~Garg, ``Badnets: Identifying vulnerabilities in the machine learning model supply chain,'' \emph{CoRR}, vol. abs/1708.06733, 2017. [Online]. Available: \url{http://arxiv.org/abs/1708.06733}
\BIBentrySTDinterwordspacing

\bibitem{DBLP:journals/corr/abs-1712-05526}
\BIBentryALTinterwordspacing
X.~Chen, C.~Liu, B.~Li, K.~Lu, and D.~Song, ``Targeted backdoor attacks on deep learning systems using data poisoning,'' \emph{CoRR}, vol. abs/1712.05526, 2017. [Online]. Available: \url{http://arxiv.org/abs/1712.05526}
\BIBentrySTDinterwordspacing

\bibitem{DBLP:conf/codaspy/ZhongLSZ020}
\BIBentryALTinterwordspacing
H.~Zhong, C.~Liao, A.~C. Squicciarini, S.~Zhu, and D.~J. Miller, ``Backdoor embedding in convolutional neural network models via invisible perturbation,'' in \emph{{CODASPY} '20: Tenth {ACM} Conference on Data and Application Security and Privacy, New Orleans, LA, USA, March 16-18, 2020}, V.~Roussev, B.~Thuraisingham, B.~Carminati, and M.~Kantarcioglu, Eds.\hskip 1em plus 0.5em minus 0.4em\relax {ACM}, 2020, pp. 97--108. [Online]. Available: \url{https://doi.org/10.1145/3374664.3375751}
\BIBentrySTDinterwordspacing

\bibitem{DBLP:conf/aaai/SahaSP20}
\BIBentryALTinterwordspacing
A.~Saha, A.~Subramanya, and H.~Pirsiavash, ``Hidden trigger backdoor attacks,'' in \emph{The Thirty-Fourth {AAAI} Conference on Artificial Intelligence, {AAAI} 2020, The Thirty-Second Innovative Applications of Artificial Intelligence Conference, {IAAI} 2020, The Tenth {AAAI} Symposium on Educational Advances in Artificial Intelligence, {EAAI} 2020, New York, NY, USA, February 7-12, 2020}.\hskip 1em plus 0.5em minus 0.4em\relax {AAAI} Press, 2020, pp. 11\,957--11\,965. [Online]. Available: \url{https://doi.org/10.1609/aaai.v34i07.6871}
\BIBentrySTDinterwordspacing

\bibitem{DBLP:conf/nips/DoanLL21}
\BIBentryALTinterwordspacing
K.~D. Doan, Y.~Lao, and P.~Li, ``Backdoor attack with imperceptible input and latent modification,'' in \emph{Advances in Neural Information Processing Systems 34: Annual Conference on Neural Information Processing Systems 2021, NeurIPS 2021, December 6-14, 2021, virtual}, M.~Ranzato, A.~Beygelzimer, Y.~N. Dauphin, P.~Liang, and J.~W. Vaughan, Eds., 2021, pp. 18\,944--18\,957. [Online]. Available: \url{https://proceedings.neurips.cc/paper/2021/hash/9d99197e2ebf03fc388d09f1e94af89b-Abstract.html}
\BIBentrySTDinterwordspacing

\bibitem{DBLP:conf/acl/KuritaMN20}
\BIBentryALTinterwordspacing
K.~Kurita, P.~Michel, and G.~Neubig, ``Weight poisoning attacks on pretrained models,'' in \emph{Proceedings of the 58th Annual Meeting of the Association for Computational Linguistics, {ACL} 2020, Online, July 5-10, 2020}, D.~Jurafsky, J.~Chai, N.~Schluter, and J.~R. Tetreault, Eds.\hskip 1em plus 0.5em minus 0.4em\relax Association for Computational Linguistics, 2020, pp. 2793--2806. [Online]. Available: \url{https://doi.org/10.18653/v1/2020.acl-main.249}
\BIBentrySTDinterwordspacing

\bibitem{DBLP:conf/ccs/LiLDZXZL21}
\BIBentryALTinterwordspacing
S.~Li, H.~Liu, T.~Dong, B.~Z.~H. Zhao, M.~Xue, H.~Zhu, and J.~Lu, ``Hidden backdoors in human-centric language models,'' in \emph{{CCS} '21: 2021 {ACM} {SIGSAC} Conference on Computer and Communications Security, Virtual Event, Republic of Korea, November 15 - 19, 2021}, Y.~Kim, J.~Kim, G.~Vigna, and E.~Shi, Eds.\hskip 1em plus 0.5em minus 0.4em\relax {ACM}, 2021, pp. 3123--3140. [Online]. Available: \url{https://doi.org/10.1145/3460120.3484576}
\BIBentrySTDinterwordspacing

\bibitem{DBLP:conf/acl/QiLCZ0WS20}
\BIBentryALTinterwordspacing
F.~Qi, M.~Li, Y.~Chen, Z.~Zhang, Z.~Liu, Y.~Wang, and M.~Sun, ``Hidden killer: Invisible textual backdoor attacks with syntactic trigger,'' in \emph{Proceedings of the 59th Annual Meeting of the Association for Computational Linguistics and the 11th International Joint Conference on Natural Language Processing, {ACL/IJCNLP} 2021, (Volume 1: Long Papers), Virtual Event, August 1-6, 2021}, C.~Zong, F.~Xia, W.~Li, and R.~Navigli, Eds.\hskip 1em plus 0.5em minus 0.4em\relax Association for Computational Linguistics, 2021, pp. 443--453. [Online]. Available: \url{https://doi.org/10.18653/v1/2021.acl-long.37}
\BIBentrySTDinterwordspacing

\bibitem{DBLP:conf/emnlp/QiCZLLS21}
\BIBentryALTinterwordspacing
F.~Qi, Y.~Chen, X.~Zhang, M.~Li, Z.~Liu, and M.~Sun, ``Mind the style of text! adversarial and backdoor attacks based on text style transfer,'' in \emph{Proceedings of the 2021 Conference on Empirical Methods in Natural Language Processing, {EMNLP} 2021, Virtual Event / Punta Cana, Dominican Republic, 7-11 November, 2021}, M.~Moens, X.~Huang, L.~Specia, and S.~W. Yih, Eds.\hskip 1em plus 0.5em minus 0.4em\relax Association for Computational Linguistics, 2021, pp. 4569--4580. [Online]. Available: \url{https://doi.org/10.18653/v1/2021.emnlp-main.374}
\BIBentrySTDinterwordspacing

\bibitem{DBLP:conf/naacl/GanL0LM00GF22}
\BIBentryALTinterwordspacing
L.~Gan, J.~Li, T.~Zhang, X.~Li, Y.~Meng, F.~Wu, Y.~Yang, S.~Guo, and C.~Fan, ``Triggerless backdoor attack for {NLP} tasks with clean labels,'' in \emph{Proceedings of the 2022 Conference of the North American Chapter of the Association for Computational Linguistics: Human Language Technologies, {NAACL} 2022, Seattle, WA, United States, July 10-15, 2022}, M.~Carpuat, M.~de~Marneffe, and I.~V.~M. Ru{\'{\i}}z, Eds.\hskip 1em plus 0.5em minus 0.4em\relax Association for Computational Linguistics, 2022, pp. 2942--2952. [Online]. Available: \url{https://doi.org/10.18653/v1/2022.naacl-main.214}
\BIBentrySTDinterwordspacing

\bibitem{DBLP:conf/acl/SunCTF0Z023}
\BIBentryALTinterwordspacing
W.~Sun, Y.~Chen, G.~Tao, C.~Fang, X.~Zhang, Q.~Zhang, and B.~Luo, ``Backdooring neural code search,'' in \emph{Proceedings of the 61st Annual Meeting of the Association for Computational Linguistics (Volume 1: Long Papers), {ACL} 2023, Toronto, Canada, July 9-14, 2023}.\hskip 1em plus 0.5em minus 0.4em\relax Association for Computational Linguistics, 2023, pp. 9692--9708. [Online]. Available: \url{https://doi.org/10.18653/v1/2023.acl-long.540}
\BIBentrySTDinterwordspacing

\end{thebibliography}
\end{document}